%% file: tmc.tex
\documentclass[lettersize,journal]{IEEEtran}
\usepackage{xspace}
\usepackage{amsmath}
\usepackage{algorithm}
\usepackage{algpseudocode}
\usepackage{xcolor}
\usepackage{graphicx}
\usepackage{subcaption}
\usepackage{booktabs}
\usepackage{pifont}
\usepackage{url}
\usepackage{eso-pic}

\algrenewcommand{\algorithmiccomment}[1]{\hfill\textcolor{black}{// #1}}

\newcommand{\semicheck}{%
  \ding{52}\rotatebox[origin=c]{-9.2}{\kern-0.7em\ding{55}}%
}

\newcommand{\blue}[1]{\textcolor{blue}{#1}}
\renewcommand{\blue}[1]{#1} %
\newcommand{\systemname}{\textit{FreeBrush}\xspace}

\title{FreeBrush: Motor Signals-Based Toothbrushing Monitoring System for Electric Toothbrushes}

\author{%
  Pinhao~Lin\IEEEauthorrefmark{1},
  Xuanyu~Liu\IEEEauthorrefmark{1}\IEEEauthorrefmark{3},
  Guoheng~Ma,
  Haoxian~Liu,
  Zongqi~Yang,
  and~Jin~Zhang\IEEEauthorrefmark{2}%
  \IEEEcompsocitemizethanks{%
    \IEEEcompsocthanksitem P. Lin, X. Liu, H. Liu, Z. Yang, and J. Zhang are with the Research Institute of Trustworthy Autonomous Systems, Department of Computer Science and Engineering, Southern University of Science and Technology, Shenzhen, China.
    \IEEEcompsocthanksitem G. Ma is with the Department of Electrical and Electronic Engineering, Southern University of Science and Technology, Shenzhen, China.
    \IEEEcompsocthanksitem \IEEEauthorrefmark{1}P. Lin and X. Liu contributed equally to this work.
    \IEEEcompsocthanksitem \IEEEauthorrefmark{3}X. Liu contributed to this work while he was a student at Southern University of Science and Technology.
    \IEEEcompsocthanksitem \IEEEauthorrefmark{2}J. Zhang is the corresponding author. E-mail: zhangj4@sustech.edu.cn
  }%
}

\begin{document}
\nocite{jin2016global,scannapieco2016oral,lockhart2012periodontal,van2011oral,yaacob2014powered,van2015efficacy,essalat2022naturalistic,chang_playful_2008,yang2024brushbuds,luo2019brush,huang2020met,chen2023lit,wang_toothfairy_2024,yangqiang_smarteeth_2025,akther_mteeth_2021,wheeler1974dental,laifen_toothbrush,usmile_y20pro_toothbrush,xiaomi_oscillation_toothbrush,mohanraj2022_bldc_review,stm32f103_price,stm32f103x8_datasheet,lks32mc08x_datasheet,krishnan_BLDC_control_theory,inceptiontime_2020,snell2017prototypical,ada_brushing_time,oral_seq_survey1_1979,oral_seq_survey2_2015,viterbi2003error,hussain2021you,huang2016toothbrushing,wang2017timeseries,rahimian2020xceptiontime,zerveas2021transformer,feng_hygiea_2024,korpela2015evaluating,ouyang2017asymmetrical,cui2024dentists,wainwright2014analysis}

\AddToShipoutPictureFG*{%
  \AtPageUpperLeft{%
    \put(48.96,-147){\parbox[t]{514.08pt}{\fontsize{8}{9}\selectfont\noindent
This article extends our workshop paper, BrushSense~\cite{lin2026brushsense}, from 6 regions to 16 tooth surfaces, adding brushing-duration estimation, stroke segmentation, improved user registration, spatio-temporal refinement, and evaluation with 18 participants. An anonymized review artifact is linked in the paper; public release is planned upon acceptance.
    }}%
  }%
  \AtPageLowerLeft{%
    \put(48.96,19){\parbox[b]{514.08pt}{\fontsize{7}{8}\selectfont\noindent
This work has been submitted to the IEEE for possible publication. Copyright may be transferred without notice, after which this version may no longer be accessible.
    }}%
  }%
}

\maketitle

\begin{abstract}
Oral health is vital for overall well-being, yet many users brush different tooth surfaces for insufficient or excessive durations. Toothbrushing monitoring systems can help improve brushing habits by reminding users of under-brushed and over-brushed surfaces, but existing approaches often rely on external sensors, increasing cost and reducing practicality. We observe that the intrinsic motor signals of electric toothbrushes can reflect the brushing surfaces and durations. When brushing different tooth surfaces, the brush head experiences different constraints from the surrounding oral cavity, changing the motor's driving currents and position feedback. Based on this observation, we propose \systemname, a toothbrushing monitoring system that uses the intrinsic motor signals of electric toothbrushes to recognize brushed tooth surfaces and estimate brushing durations without any external sensors.

We prototype the system on a commercial off-the-shelf electric toothbrush and conduct a comprehensive evaluation with 18 participants. \systemname achieves \blue{92.8\%} accuracy for 16-surface recognition and \blue{92.0\%} accuracy for brushing duration estimation. The system further shows robustness across different toothpastes, brush heads, vibration modes, usage scenarios, and long-term use, and is capable of running on edge devices. By recognizing brushed tooth surfaces and estimating the brushing duration, \systemname can help users identify under-brushed and over-brushed surfaces and build balanced brushing habits.

\end{abstract}


\input{Sections/1_intro}
\input{Sections/2_background.tex}

\input{Sections/3_insights_feas.tex}
\input{Sections/4_overview.tex}
\input{Sections/5_acquisition.tex}
\input{Sections/6_preprocessing.tex}
\input{Sections/7_surface_classification.tex}
\input{Sections/8_imp.tex}
\input{Sections/9_evaluation.tex}

\input{Sections/10_related.tex}
\input{Sections/11_discussion.tex}

\input{Sections/12_conclusion.tex}

\bibliographystyle{IEEEtran}
\bibliography{ref}

\end{document}

%% file: Sections/1_intro.tex
\section{Introduction}
Oral health is essential for overall well-being, because poor oral hygiene not only leads to dental diseases but is also associated with multiple systemic health risks~\cite{jin2016global, scannapieco2016oral, lockhart2012periodontal}. A primary factor linking poor hygiene to these health risks is the accumulation of dental plaque, which rapidly develops without effective cleaning. Daily toothbrushing serves as the most accessible and effective means of removing plaque and maintaining oral health~\cite{van2011oral, yaacob2014powered}. Recently, electric toothbrushes are widely used due to their superior plaque removal efficacy~\cite{van2015efficacy}. However, electric toothbrushes alone do not guarantee proper cleaning: many users still brush certain areas for insufficient or excessive durations, increasing the risk of localized oral diseases and gum damage~\cite{yaacob2014powered, essalat2022naturalistic}. These issues require developing toothbrushing monitoring systems that can detect under-brushed and over-brushed tooth surfaces.

Many works have developed toothbrushing monitoring systems, but they all require additional sensors or wearable devices. Video-based systems, such as Playful Toothbrush~\cite{chang_playful_2008}, use cameras to recognize brushing across 24 areas. However, they face high costs, privacy concerns, and poor performance in low light. IMU-based systems, such as BrushBuds~\cite{yang2024brushbuds} and smartwatches~\cite{luo2019brush}, use IMU sensors worn on the head or wrist to track brushing movements. However, they suffer from motion interference and require users to wear extra devices. Magnetic- and optical-based systems, such as MET~\cite{huang2020met} and LiT~\cite{chen2023lit}, use magnetic sensors or photodetectors to locate the brushed surface. However, these sensors must be installed in the environment (MET) or mounted on the toothbrush (LiT). Acoustic systems like ToothFairy~\cite{wang_toothfairy_2024} use the microphone in commodity earphones to recognize the brushed tooth. Although they need no additional sensors, they only operate when the user wears earphones. Overall, existing systems rely heavily on additional sensors or wearable devices, increasing hardware complexity and user burden (see Table~\ref{tab:comparison_property} for a detailed comparison).

\begin{table}[t]
    \caption{Comparison with representative prior works.}
    \label{tab:comparison_property}
    \centering
    \footnotesize
    \setlength{\tabcolsep}{2.5pt}
    \renewcommand{\arraystretch}{1.25}
    \begin{tabular}{@{}lccc@{}}
        \toprule
        \textbf{Work} &
        \textbf{Hardware} &
        \textbf{Usage Requirement} &
        \textbf{Resolution} \\
        \midrule
        \textbf{SmarTeeth}~\cite{yangqiang_smarteeth_2025}
            & In-Ear Mic. & Wear earphones & 16 surfaces \\
        \textbf{BrushBuds}~\cite{yang2024brushbuds}
            & Ear IMU & Wear earbuds & 6 surfaces \\
        \textbf{ToothFairy}~\cite{wang_toothfairy_2024}
            & Wired Earphones & Wear wired earphones & Tooth-by-tooth\textsuperscript{1} \\
        \textbf{LiT}~\cite{chen2023lit}
            & Photosensor & Use lighted toothbrush & 16 surfaces \\
        \textbf{MET}~\cite{huang2020met}
            & Magnetic Sensor & Set up sensor array & 15 surfaces\textsuperscript{2} \\
        \textbf{mTeeth}~\cite{akther_mteeth_2021}
            & Wrist IMU & Wear smartwatch & 9 classes \\
        \textbf{Ours}
            & None & None & 16 surfaces \\
        \bottomrule
    \end{tabular}

    \vspace{2pt}
    \begin{minipage}{\linewidth}
        \footnotesize
        \textit{Notes:} \textsuperscript{1}\,ToothFairy identifies each tooth but can hardly distinguish inner vs.\ outer surfaces;
        \textsuperscript{2}\,MET supports 15 surfaces (not the full 16).
    \end{minipage}
    \vspace{-12pt}
\end{table}

In this paper, we aim to achieve toothbrushing monitoring without any external sensors. To the best of our knowledge, we are the first to use the intrinsic motor signals of electric toothbrushes for toothbrushing monitoring. We observe that these signals can reflect the brushing surfaces and durations. When brushing different tooth surfaces, the surrounding oral cavity imposes different constraints on the brush head, disturbing its oscillation. This change is first reflected in the motor's position feedback, which tracks the brush head's motion in real time. To restore its normal oscillation, the motor adjusts its driving currents until the oscillation becomes stable again. As a result, brushing different surfaces produces distinct patterns in both the position feedback and the driving currents. From these motor signals, the system recognizes which surface is being brushed and estimates the brushing duration. In this way, the system helps users identify which surfaces are under-brushed and which are over-brushed, so that they can build balanced brushing habits.

\begin{figure*}[htbp]
    \centering
    \begin{minipage}[b]{0.30\linewidth}
        \vspace{0pt} %
        \centering
        \includegraphics[width=\linewidth]{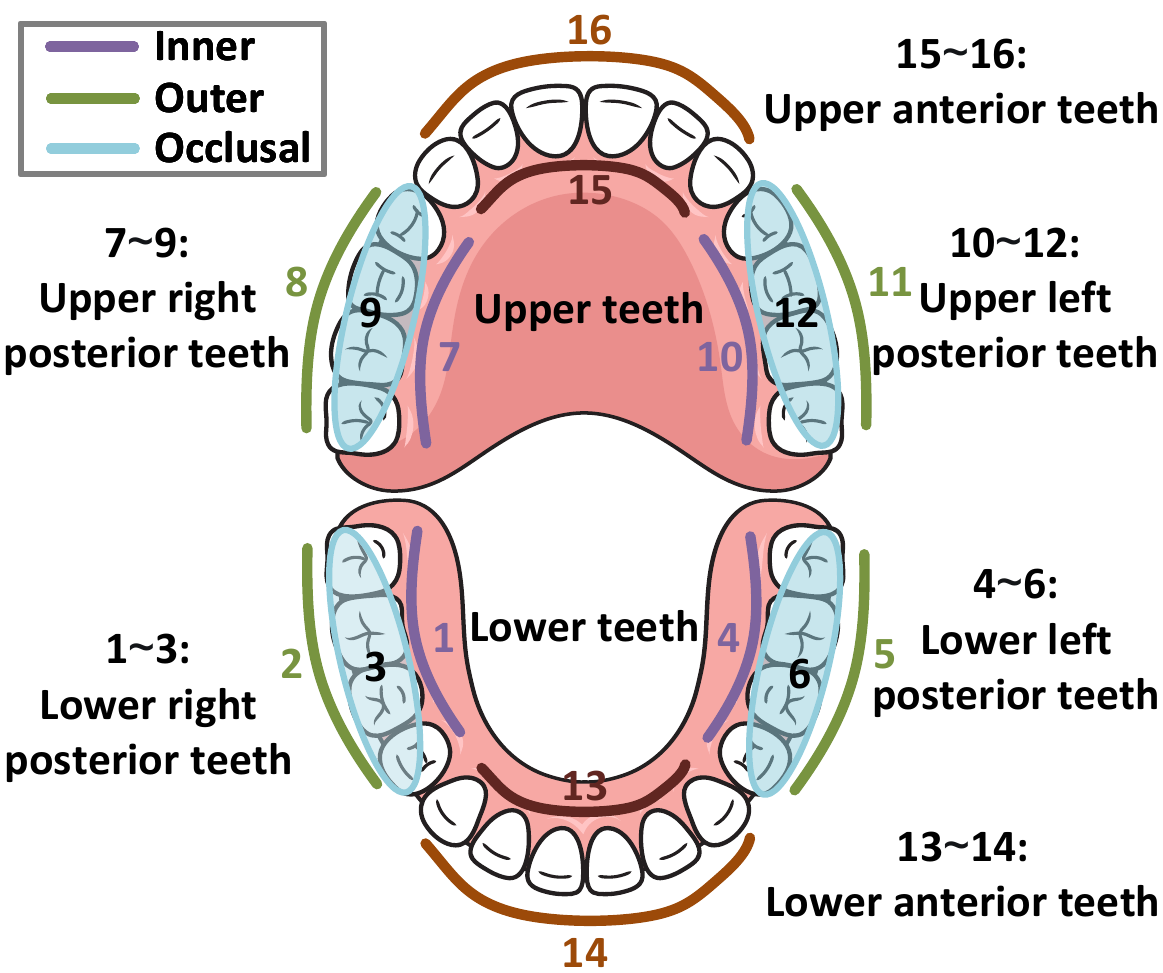}
        \caption{Sixteen-surface partition of the teeth.}
        \label{fig:oral_cavity_property}
    \end{minipage}%
    \hspace{0.02\linewidth}%
    \begin{minipage}[b]{0.35\linewidth}
        \vspace{0pt} %
        \centering
        \includegraphics[width=\linewidth]{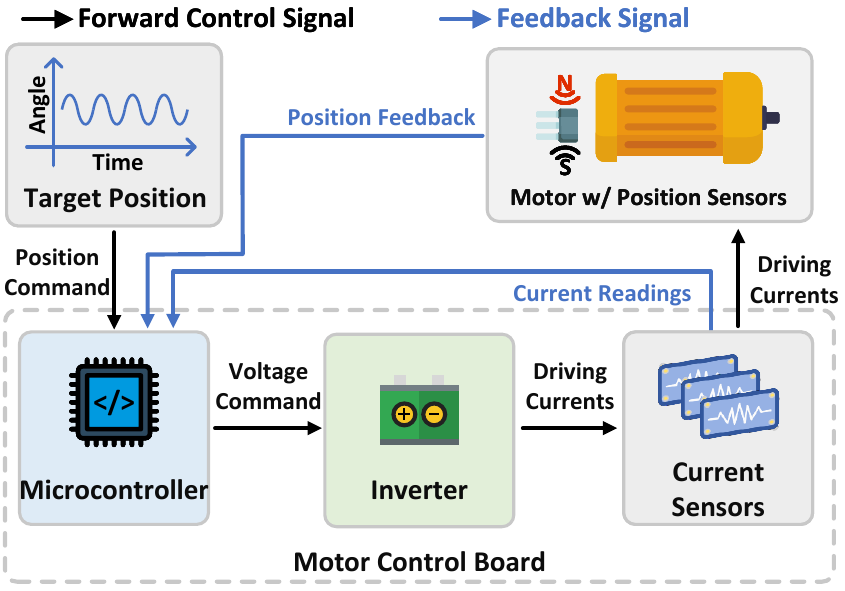}
        \caption{Main components and control loop of an electric toothbrush.}
        \label{fig:control_diagram}
    \end{minipage}%
    \hspace{0.02\linewidth}%
    \begin{minipage}[b]{0.30\linewidth}
        \vspace{0pt} %

        \vspace{-3pt}

        \centering
        \begin{subfigure}[t]{\linewidth}
            \centering
            \hspace{5pt}
            \includegraphics[width=0.68\linewidth]{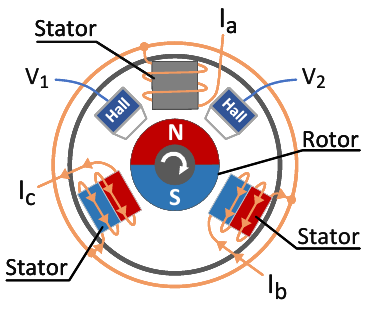}
            \vspace{-6pt} %
            \caption{Motor structure.}
            \label{fig:bldc_motor_structure_a}
        \end{subfigure}

        \vspace{2pt} %

        \begin{subfigure}[t]{\linewidth}
            \centering
            \includegraphics[width=0.7\linewidth]{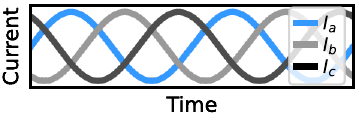}
            \vspace{-6pt} %
            \caption{Driving current waveforms.}
            \label{fig:bldc_motor_structure_b}
        \end{subfigure}

        \caption{Motor structure and driving current waveforms.}
        \label{fig:bldc_motor_structure}
    \end{minipage}
\end{figure*}

This system faces several key challenges:
\begin{itemize}
    \item Intrinsic BLDC motor signals have not been explored as a sensing modality for toothbrushing monitoring. It is unclear how brushing behaviors influence these signals and whether the signals can reflect brushing surfaces or not. To address this, we analyze how brushing affects the motor signals, and collect data on a commercial off-the-shelf (COTS) toothbrush to validate the feasibility of using this modality for surface recognition.

    \item Recognizing the brushed surface and estimating its brushing duration from motor signals is challenging. The signals have five channels and are modulated by the brush head's oscillation, which brings a heavy load for processing, and the model performs poorly when fed such raw signals directly. Also, the patterns vary across users, so a model trained on other users performs poorly on a new user. In addition, the signals are continuous and contain idle periods when the user is not brushing, and these periods would be counted into the brushing duration if not removed. To address these, we reduce the five channels to three and align them to the rotor's magnetic reference frame to remove the modulation, adapt the model to each user with a one-time registration, and segment the signals into brushing strokes to remove the idle periods.

\end{itemize}

We summarize our main contributions as follows:

\begin{itemize}

    \item We propose \systemname, the first toothbrushing monitoring system that uses the intrinsic motor signals of electric toothbrushes, without any external sensors. We analyze how the oral cavity constrains the brush head and how the constraints show in the driving currents and position feedback, and validate the feasibility on a commercial off-the-shelf toothbrush. The system recognizes the brushed surface among 16 surfaces and estimates the brushing duration on each surface, helping users find under-brushed and over-brushed surfaces. Review copy: \url{https://anonymous.4open.science/r/FreeBrush_Artifact/}.

    \item We design a lightweight surface recognition and duration estimation algorithm for the motor signals. It aligns the signals to the rotor's magnetic reference frame to remove the modulation from the brush head's oscillation, segments brushing into individual strokes and removes the idle periods, adapts the model to each user with a one-time registration, and refines the stroke predictions using the oral anatomy and the continuity of brushing.

    \item We build a prototype on a commercial electric toothbrush and conduct a comprehensive evaluation with 18 participants. The system achieves \blue{92.8\%} accuracy for 16-surface recognition and \blue{92.0\%} accuracy for brushing duration estimation. It stays robust across different toothpaste types, brush heads and their wear, vibration modes, usage scenarios, and long-term use. It is also capable of running on edge devices.

\end{itemize}

%% file: Sections/2_background.tex
\section{Background and Motivation}
\label{sec:background}
In this section, we explain why intrinsic motor signals can be used for toothbrushing monitoring. We first provide background on how these signals vary across brushing surfaces from the perspectives of oral cavity structure, electric toothbrush's motor structure and control. Then, we present a feasibility study that validates this idea.

\subsection{Structure of the Oral Cavity}
\label{sec:background.oral_cavity}

Toothbrushing monitoring primarily measures how long each surface of each tooth is brushed, ensuring each surface is adequately cleaned. Therefore, we partition the mouth into 16 tooth surfaces as illustrated in Fig.~\ref{fig:oral_cavity_property}, following conventions in prior work~\cite{chen2023lit, yangqiang_smarteeth_2025}. These surfaces are defined based on tooth row (lower vs.\ upper), side (right vs.\ left), and surface type (outer, inner, and occlusal). Our primary goal in this work is to determine which tooth surface is being brushed at each stroke and to accumulate the brushing time on each surface, which then allows the system to identify under-brushed and over-brushed surfaces.

Different tooth surfaces face different surrounding structures~\cite{chen2023lit}. The teeth are surrounded by the lips, cheeks, palate, tongue, and gingiva (Fig.~\ref{fig:oral_cavity_property}). The front teeth are arranged horizontally at the front of the mouth, and the molars lie behind them along both sides. The outer surfaces of the front teeth face the lips, while the outer surfaces of the molars face the cheeks. The inner surfaces of the upper teeth face the palate, while the inner surfaces of the lower teeth face the tongue. The occlusal surfaces face up or down, perpendicular to the inner and outer surfaces.

The oral cavity constrains the brush head differently when brushing different surfaces, so the motor signals differ from surface to surface. When brushing the outer surfaces of the molars, the cheek presses the brush head firmly from the side; on the inner upper surfaces, the palate presses it from above; on the inner lower surfaces, the tongue presses it more softly from below. These constraints differ in direction and strength, so they disturb the oscillation differently. The occlusal surfaces and the front teeth receive little constraint, as almost nothing presses on the occlusal surfaces and the lips press only lightly on the front teeth. Their signals differ instead through bristle contact, as the occlusal surfaces are wide and uneven, covered with pits and cusps, while the front teeth are narrow with smooth outer surfaces~\cite{wheeler1974dental}. To keep the brush head oscillating steadily under these disturbances, the motor adjusts its driving currents (Sec.~\ref{sec:background.bldc_structure}), and the adjustments leave a distinct pattern for each surface. From these patterns, the system recognizes which surface is being brushed.

\subsection{Electric Toothbrush Structure and Control Mechanism}
\label{sec:background.bldc_structure}

Our system recognizes brushing surfaces from the electric toothbrush's intrinsic motor signals, so it is important to understand why these signals change with the brushed surface. This requires knowledge of the toothbrush's structure and control mechanism, which we introduce in this subsection.

An electric toothbrush mainly consists of a motor and a control board that drives it, as shown in Fig.~\ref{fig:control_diagram}. The motor is typically a Brushless Direct Current (BLDC) motor, which is becoming the mainstream choice in electric toothbrushes~\cite{laifen_toothbrush, usmile_y20pro_toothbrush, xiaomi_oscillation_toothbrush}. As shown in Fig.~\ref{fig:bldc_motor_structure}(a), the motor has two main parts. The rotor carries permanent magnets and is firmly connected to the brush head, so the brush head moves with it. The stator is fixed around the rotor and has three windings arranged \(120^\circ\) apart~\cite{mohanraj2022_bldc_review}. The motor also carries two sensors that track the rotor's position (\(V_1\) and \(V_2\) in the figure), whose role we explain below.

The toothbrush's control board makes the brush head oscillate by driving currents through the windings. Its microcontroller receives a target position command that specifies how the brush head should oscillate, and computes voltage commands, which an inverter converts into voltages on the windings. The resulting driving currents \(\{I_a, I_b, I_c\}\)\footnote{In BLDC control theory, the currents driving the motor are usually called ``phase currents"; we use ``driving currents" for clarity.} generate a magnetic field that pulls the rotor. To keep pulling the rotor as it turns, the magnetic field must stay ahead of the rotor's own magnetic field. The driving currents therefore have to change with the rotor's position. Fig.~\ref{fig:bldc_motor_structure}(b) shows an ideal example of the three driving currents during rotor rotation under no load.

To keep the brush head oscillating steadily during brushing, the toothbrush's control board continuously tracks the rotor's position and adjusts the driving currents. The currents cannot be pre-defined, as the rotor's initial position is unknown and the oral cavity keeps disturbing its oscillation. With a pre-defined current sequence, the magnetic field would fall out of step with the rotor, could no longer pull it, and the brush head would stall. Motors therefore include position sensors\footnote{In BLDC control theory, the sensors to track the rotor's location are usually called ``position encoders"; we use ``position sensors" for clarity.} to track the rotor's rotation. Electric toothbrushes commonly use two Hall sensors, offset by \(90^\circ\), and each outputs a voltage proportional to the rotor's magnetic field strength. The two voltages form the position feedback signal \(\{V_1, V_2\}\), which the board reads for fine-grained adjustment.

Besides the rotor's position, the toothbrush's control board also needs the actual driving currents. It can only regulate the voltage applied to the windings through the inverter, but the magnetic field that pulls the rotor is set by the currents (Amp\`ere's law). If it regulated the voltage alone, the currents would drift with the rotor's speed and the brushing load, the magnetic field would drift with them, and the brush head could not hold its target oscillation. The board therefore includes current sensors that measure the actual driving currents \(\{I_a, I_b, I_c\}\) and adjusts the voltage in a closed loop until the currents reach their targets (Fig.~\ref{fig:control_diagram}). We refer to the driving currents and the position feedback together as the motor signals, which serve as the input to our system.

These sensors are required by BLDC motor control itself, not added for our system. To regulate the motor's motion, the toothbrush's control board must track the rotor's position and the actual driving currents~\cite{mohanraj2022_bldc_review}, so the position feedback and driving currents come with the motor rather than with our system.

When brushing different surfaces, the oral cavity disturbs the brush head's oscillation differently and leaves a distinct pattern in the motor signals. Depending on the surface, the surrounding oral tissue or the tooth itself pushes against the brush head. The push shifts the oscillation off its center, and the contact damps its amplitude. These two effects differ in direction and strength across surfaces, as the surrounding structures differ. The cheek presses the brush head firmly from the side, while the tongue pushes it softly from below (Sec.~\ref{sec:background.oral_cavity}). The rotor therefore deviates from its target motion in a surface-specific way. The deviation shows up directly in the position feedback, which tracks the rotor in real time. The toothbrush's control board counters the deviation by adjusting the driving currents until the rotor returns to its target motion, so the correction shows up in the driving currents. The deviation and the correction both depend on the brushed surface, and the system recognizes the surface from these two signals.

\begin{figure*}[htbp]
    \centering
    \begin{subfigure}[b]{0.24\linewidth}
        \centering
        \includegraphics[width=\linewidth]{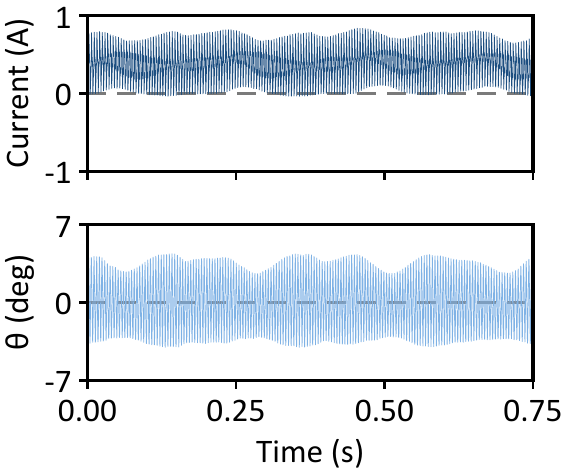}
        \caption{Upper-left outside}
    \end{subfigure}
    \hfill
    \begin{subfigure}[b]{0.24\linewidth}
        \centering
        \includegraphics[width=\linewidth]{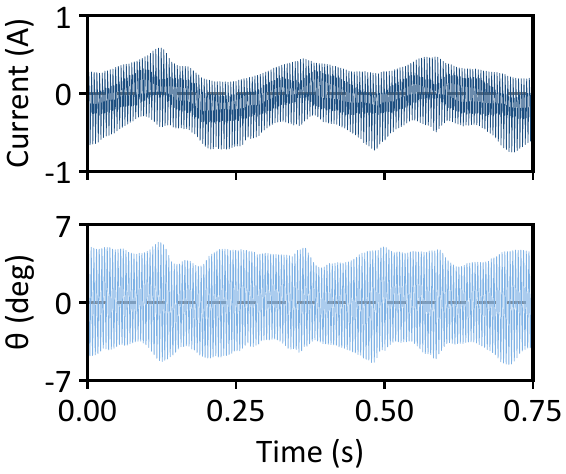}
        \caption{Upper-left inside}
    \end{subfigure}
    \hfill
    \begin{subfigure}[b]{0.24\linewidth}
        \centering
        \includegraphics[width=\linewidth]{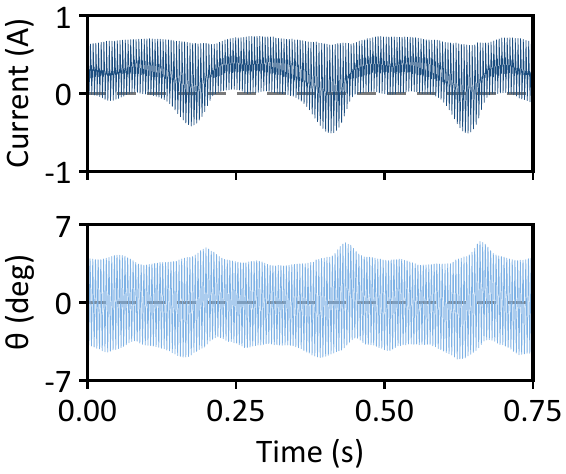}
        \caption{Lower-left inside}
    \end{subfigure}
    \hfill
    \begin{subfigure}[b]{0.24\linewidth}
        \centering
        \includegraphics[width=\linewidth]{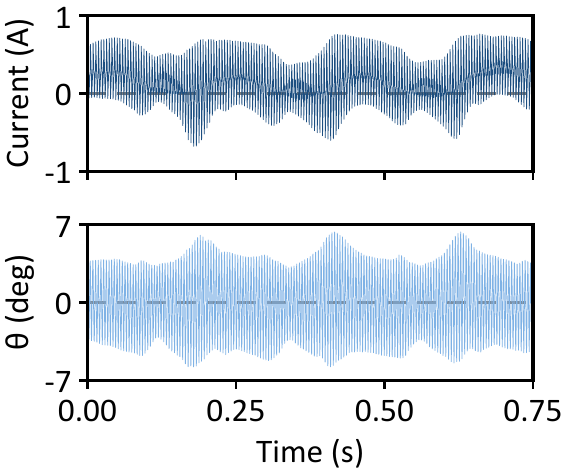}
        \caption{Upper-right inside}
    \end{subfigure}
\caption{One channel of the driving currents (top, dark blue) and position feedback $\theta$ (bottom, light blue) recorded on four tooth surfaces within a single brushing session: (a) upper-left outside, (b) upper-left inside, (c) lower-left inside, and (d) upper-right inside. Dashed lines mark zero.}
\label{fig:feasibility_example}
\end{figure*}

\begin{figure*}[t]
    \centering
    \includegraphics[width=0.9\linewidth]{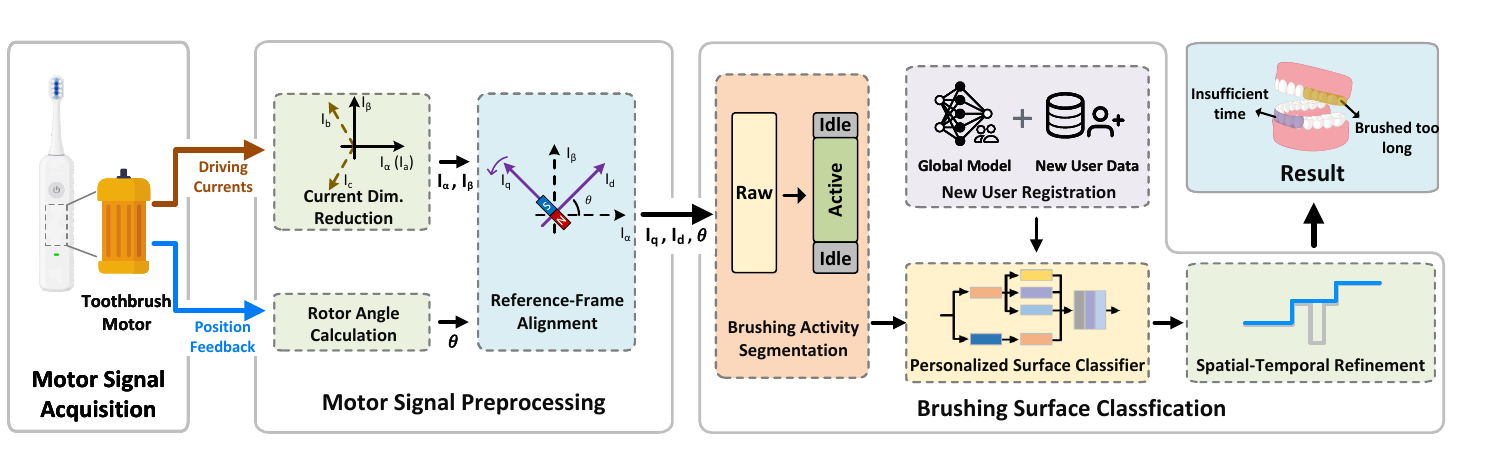}
    \caption{System overview of \systemname.}
    \label{fig:system_overview}
\end{figure*}

\subsection{Feasibility Study}
The previous analysis shows that brushing different surfaces leaves distinct patterns in the motor signals, making it possible to recognize the brushed surface from these signals. \textit{But do the intrinsic motor signals of a real toothbrush exhibit such patterns?} To verify this, we conduct a preliminary feasibility study. We collect motor signals from different tooth surfaces using the hardware prototype based on a commercial electric toothbrush, as described in Sec.~\ref{sec:imp.hardware_prototype}. Fig.~\ref{fig:feasibility_example} shows one channel of the driving currents and the position feedback after preprocessing (Sec.~\ref{sec:preprocessing}), recorded on four representative surfaces within a single brushing session.

The recorded signals show that brushing different surfaces leaves distinct patterns, even between very similar surfaces. On the upper-left outside surface (Fig.~\ref{fig:feasibility_example}(a)), the cheek wraps the brush head. The band stays above zero and is the narrowest of the four. On the upper-left inside surface (b), little constrains the brush head, with only the palate above it. The band is uneven with large peaks and lies mostly below zero. On the lower-left inside surface (c), the tongue touches the brush head from below but does not wrap it. The band is smooth and regular, with sharp peaks. On the upper-right inside surface (d), the palate is again the only structure above the brush head. The band is uneven once more, but lies mostly above zero. The band is steadiest on (a), where the brush head is most constrained, and most uneven on (b) and (d), where it is least constrained. The patterns therefore differ between the two surfaces of the same tooth ((a) and (b)), between the upper and lower surfaces on the same side ((b) and (c)), between the left and right surfaces in the same row ((b) and (d)), and between two centrally symmetric surfaces ((c) and (d)). These differences follow how the brush head is constrained on each surface (Sec.~\ref{sec:background.oral_cavity}), supporting the feasibility of using intrinsic motor signals for toothbrushing monitoring.

%% file: Sections/4_overview.tex
 \section{System Overview}

Our aim is to use motor signals from electric toothbrushes for toothbrushing monitoring without external sensors. The system recognizes the brushed surface and estimates the brushing duration on each surface, helping users build balanced brushing habits. However, the motor signals have five channels, which brings a heavy load for processing, and they are modulated by the brush head's oscillation, which hides the surface-related patterns. As a result, a lightweight model performs poorly when fed such raw signals directly. We therefore reduce the data dimension to ease the processing load, and align the currents to the rotor's magnetic reference frame to remove the modulation and reveal the surface-related patterns. Moreover, the signals form a continuous stream interleaved with idle periods, and these periods would be counted into the brushing duration if not removed. We therefore segment the stream into individual brushing strokes and discard the idle periods. As the brush head stays on a single surface within each stroke, each segment corresponds to exactly one surface, and non-brushing time is excluded from duration estimation. Additionally, the oral structures differ across users, such as tooth size and the tightness of the surrounding soft tissue, so the surface-related patterns can slightly differ across users, degrading the classification accuracy. We therefore adapt the classifier to each user through a one-time registration. Finally, occasional misclassifications may occur within otherwise correct sequences, and we suppress these short isolated errors by exploiting the oral anatomy and the continuity of brushing.

We design \systemname, as shown in Fig.~\ref{fig:system_overview}, consisting of three modules:
\textbf{(i) Motor Signal Acquisition.} We capture the intrinsic motor signals, including the three driving currents and the position feedback, from COTS BLDC toothbrushes without interfering with their normal operation.
\textbf{(ii) Motor Signal Pre-processing.} The collected signals are first reduced from five channels to three to ease the processing load. We then align the signals to the rotor's magnetic reference frame to remove the fluctuations caused by the brush head's oscillation and reveal the surface-related patterns.
\textbf{(iii) Surface Recognition and Duration Estimation.} The preprocessed signals are segmented into individual brushing strokes so that each segment corresponds to a single tooth surface, and idle periods are removed to exclude non-brushing time from duration estimation. We then apply a lightweight InceptionTime-based model to identify the brushing surface of each segment. To handle the differences across users, a global model is first pretrained on multi-user data and then adapted with a few samples collected in a one-time registration. The registration samples are also used to construct user-specific surface reference vectors. Finally, the predictions are refined using the oral anatomy and the continuity of brushing, and the brushing duration on each surface is calculated by accumulating the lengths of the corresponding segments.

%% file: Sections/5_acquisition.tex
\section{Motor Signal Acquisition}
\label{sec:motor_signal_acquisition}

The input to \systemname\ is five channels of intrinsic motor signals, which together record both how the oral cavity disturbs the brush head's oscillation and how the motor counters the disturbance. Three channels are the driving currents $\{I_a, I_b, I_c\}$, the currents that the control board applies through the motor's three stator windings to keep the brush head oscillating. The control board adjusts these currents to counter the disturbances from the oral cavity, so they carry surface-related patterns (Sec.~\ref{sec:background.bldc_structure}). The other two channels are the position feedback $\{V_1, V_2\}$, the voltages from the two built-in Hall sensors that track the rotor's position, which record how the oscillation deviates under these disturbances. All five channels are sampled at 10~kHz.

Acquiring these signals requires no external sensors, as all five channels already exist on the toothbrush's motor control board. To keep the brush head oscillating steadily, the control board must measure the actual driving currents and track the rotor's position for its closed-loop control (Sec.~\ref{sec:background.bldc_structure}), so the five channels come with motor control itself rather than with our system. The 10~kHz sampling rate is also easy to reach. Even an entry-level microcontroller such as the STM32F103, which costs less than \blue{\$2}~\cite{stm32f103_price}, samples a single channel at up to \blue{1~MHz}~\cite{stm32f103x8_datasheet}. The microcontroller on a toothbrush's control board is generally far more capable, as it also has to regulate the motor in real time~\cite{lks32mc08x_datasheet}. Manufacturers could therefore support our sensing mechanism natively, and integrating this system into future designs would incur virtually no extra cost.

%% file: Sections/6_preprocessing.tex
\section{Motor Signal Preprocessing}
\label{sec:preprocessing}
The raw signals from the toothbrush are three driving currents and two position-feedback voltages. We first reduce the three currents to two channels, then align these channels to the rotor's magnetic reference frame.

\begin{figure}[t]
    \centering
    \begin{subfigure}[b]{\linewidth}
        \centering
        \includegraphics[width=0.8\linewidth]{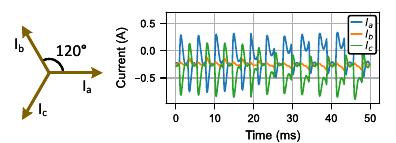}
        \caption{Raw three-phase driving currents $(I_a, I_b, I_c)$ with a $120^{\circ}$ phase difference.}
        \label{fig:three_phase_raw}
    \end{subfigure}

    \vspace{4pt}
    \begin{subfigure}[b]{\linewidth}
        \centering
        \includegraphics[width=0.8\linewidth]{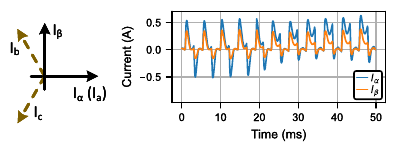}
        \caption{Projection onto the orthogonal $\alpha$-$\beta$ plane after Clarke Transformation.}
        \label{fig:clarke_transform}
    \end{subfigure}

    \vspace{4pt}
    \begin{subfigure}[b]{\linewidth}
        \centering
        \includegraphics[width=0.8\linewidth]{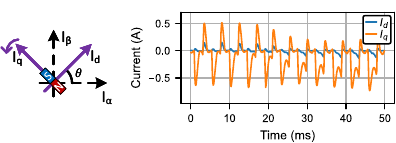}
        \caption{Alignment to the rotating $d$-$q$ frame using Park Transformation.}
        \label{fig:park_transform}
    \end{subfigure}

    \vspace{-5pt}
    \caption{Illustration of current dimension reduction and reference-frame alignment.}
    \label{fig:clarke_park_transformations}
    \vspace{-12pt}
\end{figure}

\subsection{Current Dimension Reduction}
The three driving currents can be represented by two channels without losing information. In a balanced BLDC motor system, the three currents $\{I_a, I_b, I_c\}$ have a $120^{\circ}$ phase shift and satisfy $I_a + I_b + I_c = 0$, so they lie in an orthogonal $\alpha$-$\beta$ plane. We move them into this plane with the Clarke Transformation~\cite{krishnan_BLDC_control_theory}, which lowers the computation cost of the later stages. The transformation is defined as:
\begin{equation}
\begin{bmatrix}
I_\alpha \\
I_\beta
\end{bmatrix}
=
\begin{bmatrix}
1 & -\frac{1}{2} & -\frac{1}{2} \\
0 & \frac{\sqrt{3}}{2} & -\frac{\sqrt{3}}{2}
\end{bmatrix}
\begin{bmatrix}
I_a \\
I_b \\
I_c
\end{bmatrix}.
\end{equation}
Fig.~\ref{fig:clarke_park_transformations}(a) shows the raw three-phase driving currents, and Fig.~\ref{fig:clarke_park_transformations}(b) shows the two orthogonal components $(I_\alpha, I_\beta)$ after the Clarke Transformation.

\subsection{Reference-Frame Alignment}

The driving currents of a BLDC motor vary not only with brushing-induced torque but also with the rotor's magnetic orientation, introducing angle-dependent modulation into the signals. After the \textit{Current Dimension Reduction}, this modulation persists in the $\alpha$-$\beta$ currents, causing the two channels to exhibit similar periodic patterns. To separate this effect and uncover brushing dynamics, we align the current signals to the rotor's magnetic reference frame using the Park Transformation~\cite{krishnan_BLDC_control_theory}, as illustrated in Fig.~\ref{fig:clarke_park_transformations}(c). The transformation is defined by:
\begin{equation}
\label{eq:park_result}
\begin{bmatrix}
I_d \\[3pt]
I_q
\end{bmatrix}
=
\begin{bmatrix}
\cos(\theta) & \sin(\theta) \\[3pt]
-\sin(\theta) & \cos(\theta)
\end{bmatrix}
\begin{bmatrix}
I_\alpha \\[3pt]
I_\beta
\end{bmatrix},
\end{equation}
where $\theta$ is the rotor angle. We obtain $\theta$ from the position feedback $\{V_1, V_2\}$, which comes from two Hall sensors placed $90^\circ$ apart (Sec.~\ref{sec:background.bldc_structure}, Fig.~\ref{fig:bldc_motor_structure}(a)) and therefore follows the rotor's magnetic field along two orthogonal directions:
\begin{equation}
\theta = \operatorname{atan2}(V_2, V_1).
\end{equation}

The $d$-$q$ frame separates the current that drives the brush head from the current that builds the motor's magnetic field. $I_q$ is the driving part, so it carries the brushing signal. When the brush head meets more resistance, the motor needs more force to keep its oscillation, and the driver supplies this force by raising $I_q$. $I_d$ is the part that builds the magnetic field, and it changes when the driver adjusts this field under the load. Because the frame turns with the rotor, both components are now free of the modulation caused by the rotor's magnetic orientation. $I_q$ still swings back and forth with the brush head's oscillation, as the motor drives the head in one direction and then the other, but this swing now follows the load on the brush head rather than the rotor's magnetic orientation. We then downsample $\{I_d, I_q, \theta\}$ by a factor of 25, from the 10~kHz acquisition rate to 400~Hz, to reduce the computation of the later stages.

%% file: Sections/7_surface_classification.tex
\section{Surface Recognition and Duration Estimation}
We introduce the classification algorithm in \systemname, which estimates the brushing duration on each of the 16 surfaces from the preprocessed motor signals. It segments the signals into brushing strokes, classifies the surface of each stroke with a classifier personalized by a one-time registration, refines the predictions along the brushing sequence, and accumulates the brushing time on each surface.

\begin{algorithm}[t]
  \caption{Brushing Stroke Segmentation}
  \label{alg:idle_segmentation}
  \begin{algorithmic}[1]
    \Require Transformed driving currents $I_d, I_q$ of length $N$; minimum durations $T_{\text{idle}}, T_{\text{seg}}$
    \Ensure Brushing segments $\mathcal{S}$

    \State $env_d \gets \textsc{UpperEnvelope}(I_d)$
    \State $env_q \gets \textsc{UpperEnvelope}(I_q)$

    \algorithmiccomment{Extract upper envelopes}

    \State $(m_d,m_q,\tau_d,\tau_q) \gets \textsc{AutoCalibration}$

    \algorithmiccomment{Precomputed during one-time device calibration}

    \State $is\_idle[t]\!\gets\!(|env_d[t]-m_d|\!<\!\tau_d)\land(|env_q[t]-m_q|\!<\!\tau_q)$

    \algorithmiccomment{Samples within baseline thresholds}

    \State $\mathcal{B}\gets\{[b^s,b^e]\!\mid\!is\_idle[t]{=}1,\ \forall t\!\in\![b^s,b^e],\ b^e{-}b^s{\ge}T_{\text{idle}}\}$

    \algorithmiccomment{Idle intervals lasting at least $T_{\text{idle}}$}

    \State $\mathcal{S}_{raw} \gets \text{complement of } \mathcal{B}$

    \algorithmiccomment{Active regions between idle intervals}

    \State $\mathcal{S} \gets \{[s,e]\!\in\!\mathcal{S}_{raw}
           \mid e{-}s\!\ge\!T_{\text{seg}}\}$

    \algorithmiccomment{Filter short segments}

    \Return $\mathcal{S}$
  \end{algorithmic}
\end{algorithm}

\subsection{Brushing Stroke Segmentation}
\label{sec:segmentation}
\systemname divides the signals into brushing segments, each containing one stroke on a tooth surface, and removes the idle periods in between to avoid counting non-brushing time in duration estimation and reduce unnecessary computation. The system infers the idle motor state directly from the driving currents. When the brush head is not touching the teeth, both $I_d$ and $I_q$ remain close to their baselines with limited variation. Once the bristles touch a tooth surface, the surface and the surrounding soft tissue load the brush head, and both currents deviate from their baselines and fluctuate.

To minimize the influence of oscillation on the driving currents, idle detection is performed on the upper envelopes of $I_d$ and $I_q$ (derived from Eq.~\ref{eq:park_result}). In the following, we consider only the upper envelopes for simplicity. Each device performs a one-time automatic calibration on the envelopes to determine the baseline levels $m_d,m_q$ and the stability thresholds $\tau_d,\tau_q$. The baselines are the medians of $env_d$ and $env_q$, and the thresholds are 3 times the median absolute deviations from these baselines. The thresholds are thus wider than the no-contact fluctuation, so that a small drift during an idle period does not break it into separate segments.

Given the envelopes $env_d$ and $env_q$, the motor state at time $t$ is classified as idle if both conditions are satisfied:
\begin{equation}
|env_d[t] - m_d| < \tau_d, \quad |env_q[t] - m_q| < \tau_q.
\end{equation}

\begin{figure}[t]
    \centering
    \includegraphics[width=1.0\linewidth]{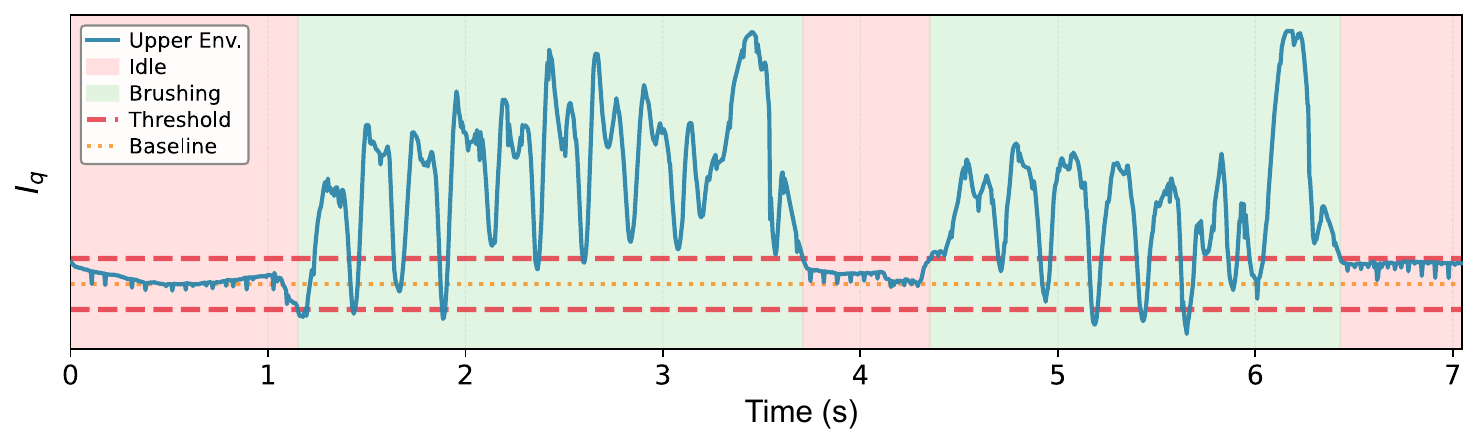}
    \caption{Illustration of the brushing stroke segmentation using $I_q$.}
    \label{fig:segmentation_example}
    \vspace{-8pt}
\end{figure}

To avoid misclassifications due to transient fluctuations, an idle state must persist for at least $T_{\text{idle}}$ consecutive samples before a transition is confirmed, and a brushing segment is the interval between two idle periods that exceeds the minimum duration $T_{\text{seg}}$. Algorithm~\ref{alg:idle_segmentation} summarizes the procedure. Empirically, we set $T_{\text{idle}}$ to 0.2 s and $T_{\text{seg}}$ to 0.8 s. Fig.~\ref{fig:segmentation_example} shows a segmentation result using $I_q$. Each segment covers one stroke, and the idle periods between the strokes are excluded from the following steps.

\subsection{Brushing Surface Classifier}
\label{sec:classification.classifier}
The classifier takes segments of the preprocessed driving currents and rotor angle $\{I_d, I_q, \theta\}$ as input and predicts the brushed surface for each segment. Each segment is Z-score normalized. Since segments vary in length, we pad them all to a fixed length of $L=3200$ samples, i.e., 8~s at 400~Hz, which is longer than any segment in our data.

\begin{figure}[t]
    \centering
    \includegraphics[width=\linewidth]{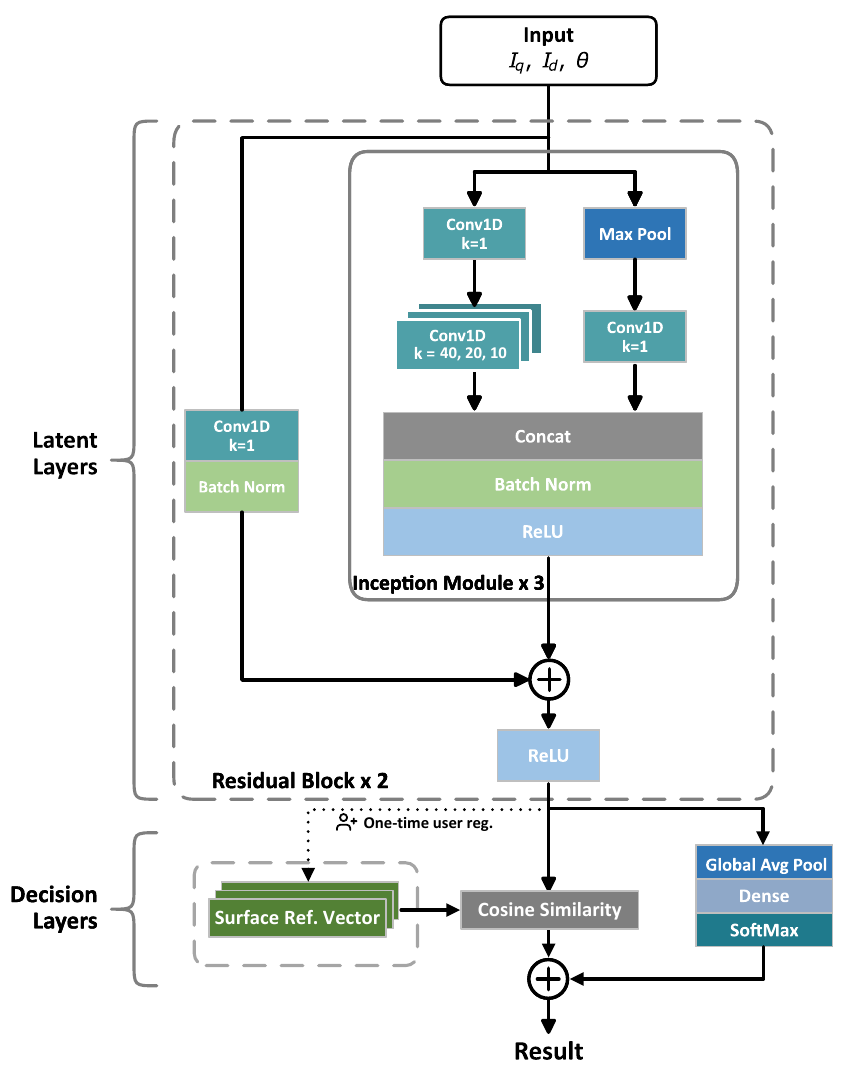}
    \caption{Architecture of \textit{Brushing Surface Classifier}.}
    \label{fig:inception_registration}
    \vspace{-8pt}
\end{figure}

We design the \textit{Brushing Surface Classifier} based on InceptionTime \cite{inceptiontime_2020}, as shown in Fig.~\ref{fig:inception_registration}. The model has latent layers and decision layers. The latent layers follow InceptionTime and extract features from the input signals, where each Inception module applies convolutional kernels of three lengths (40, 20, and 10 samples) in parallel to capture multiple time scales. We refer readers to the work of Ismail Fawaz et al. \cite{inceptiontime_2020} for the details. The decision layers map these features to surface predictions, combining the classification head of InceptionTime with a registration branch.

The registration branch scores each segment by its similarity to the user's own registration samples. This follows the prototype-based design in few-shot learning \cite{snell2017prototypical}, where each class is summarized as the average feature of its samples and inputs are classified by similarity to these averages. In our case, each surface is summarized as a surface reference vector, generated during registration (Sec.~\ref{sec:classification.reg}).

We summarize each segment into a fixed-length registration feature $\mathbf{z}$, and stack the user's 16 surface reference vectors into a reference matrix $\mathbf{R}$. For one segment, the latent layers produce a feature map $\mathbf{H}$ of size $C\times T$, where $C=128$ is the number of channels and $T$ is the number of time steps. We reduce the time axis by taking the mean and the maximum of each channel:
\begin{equation}
    \boldsymbol{\mu}=\operatorname{Mean}_{t}(\mathbf{H}),\qquad
    \mathbf{m}=\operatorname{Max}_{t}(\mathbf{H}),
\end{equation}
where $\boldsymbol{\mu}$ gives the average response of each channel over the segment and $\mathbf{m}$ gives its strongest response. We concatenate the two vectors and apply L2 normalization to obtain the registration feature $\mathbf{z}$, a vector of length $2C$:
\begin{equation}
    \mathbf{z}=\frac{[\boldsymbol{\mu};\mathbf{m}]}{\lVert[\boldsymbol{\mu};\mathbf{m}]\rVert_2}.
\end{equation}
The normalization removes the overall magnitude, so the comparison depends on the relative pattern of the channel responses rather than their strength. The reference matrix $\mathbf{R}$ is a $16\times 2C$ matrix that stacks the 16 surface reference vectors, each summarizing the user's registration samples on one surface. We introduce how they are generated in the next subsection (Sec.~\ref{sec:classification.reg}).

The branch scores a test segment by comparing its feature $\mathbf{z}$ with the reference vector of each surface, and the surface whose reference is most similar to $\mathbf{z}$ receives the highest score. The product $\mathbf{z}\mathbf{R}^{\mathsf T}$ computes the 16 comparisons at once, and each entry is the cosine similarity between the segment and one surface, as both $\mathbf{z}$ and the reference vectors have unit length.

The two branches are then combined into one prediction. We scale the similarities by a factor $s$ before the softmax, as they lie in $[-1,1]$ while the logits $\mathbf{l}$ from the classification head are unbounded. The scale is estimated from the registration data as the ratio of the standard deviations of the two, so it requires no manual tuning. The final probability averages the two predictions:
\begin{equation}
    \label{eq:fusion}
    \mathbf{p}=0.5\,\operatorname{softmax}(\mathbf{l})+0.5\,\operatorname{softmax}(s\,\mathbf{z}\mathbf{R}^{\mathsf T}).
\end{equation}
The classification head keeps the shared patterns learned from multiple users, while the registration branch adapts the predictions to the new user.

\subsection{New-user Registration}
\label{sec:classification.reg}
\systemname adapts the classifier to each new user through a one-time registration. Each user has unique oral structures, so the surface-related patterns vary across users, and a global model trained on multiple users may not generalize well to a new user. Registration is common in toothbrushing monitoring \cite{yangqiang_smarteeth_2025,yang2024brushbuds,wang_toothfairy_2024,akther_mteeth_2021}, and we adopt a protocol similar to SmarTeeth \cite{yangqiang_smarteeth_2025}. In a registration session, an animation indicates the 16 surfaces in a preset order and moves to the next surface after 8 seconds. One session takes about 2 minutes, which aligns with the recommendation of the American Dental Association (ADA)~\cite{ada_brushing_time}. As the user follows the animation strictly, the indicated sequence serves as the labels of the collected data. Users can complete multiple sessions in their daily brushing routines to further improve the accuracy.

A few registration samples are sufficient to personalize the model. Trained on multi-user data, the global model learns the patterns that distinguish the surfaces and are shared across users, while the differences across users, such as tooth size and the tightness of the surrounding soft tissue, cause user-specific deviations from these shared patterns. The registration therefore only needs to adapt the model to these deviations, instead of learning to distinguish the surfaces from scratch.

We personalize the model in two ways, fine-tuning and reference vector generation. Fine-tuning adapts the weights of the global model to the new user, and the surface reference vectors let the registration branch compare each test segment with the user's own samples. Both use the registration samples, which go through the same preprocessing and segmentation algorithm as the training data.

\textbf{Fine-tuning.}
Fine-tuning continues training the global model on the registration samples, so the network weights shift toward the new user's signals. It updates both the latent layers and the classification head, using 80\% of the pretraining epochs and a learning rate reduced to 1/10. The small learning rate keeps the personalized model close to the global one, so it retains the shared patterns while adapting to the new user.

\textbf{Reference Vector Generation.}
The surface reference vectors bring the user's own samples into the decision at test time, as fine-tuning moves the weights only slightly and the classification head still decides mostly by the patterns shared across users. Each vector is built from this user's samples alone, and it adds no trainable layer. We pass the registration samples through the latent layers and compute the registration feature of each sample. For surface $k$, we average the registration features of its samples and normalize the result to obtain its surface reference vector $\mathbf{r}_k$:
\begin{equation}
    \mathbf{r}_k=\frac{\frac{1}{N_k}\sum_{i:y_i=k}\mathbf{z}_i}{\left\lVert\frac{1}{N_k}\sum_{i:y_i=k}\mathbf{z}_i\right\rVert_2},
\end{equation}
where $\mathbf{z}_i$ is the registration feature of sample $i$, $y_i$ is its surface label, and $N_k$ is the number of samples labeled $k$. Stacking the 16 surface reference vectors gives the reference matrix $\mathbf{R}$ used by the registration branch in Eq.~\ref{eq:fusion}.

\begin{figure}[t]
    \centering
    \includegraphics[width=0.9\linewidth]{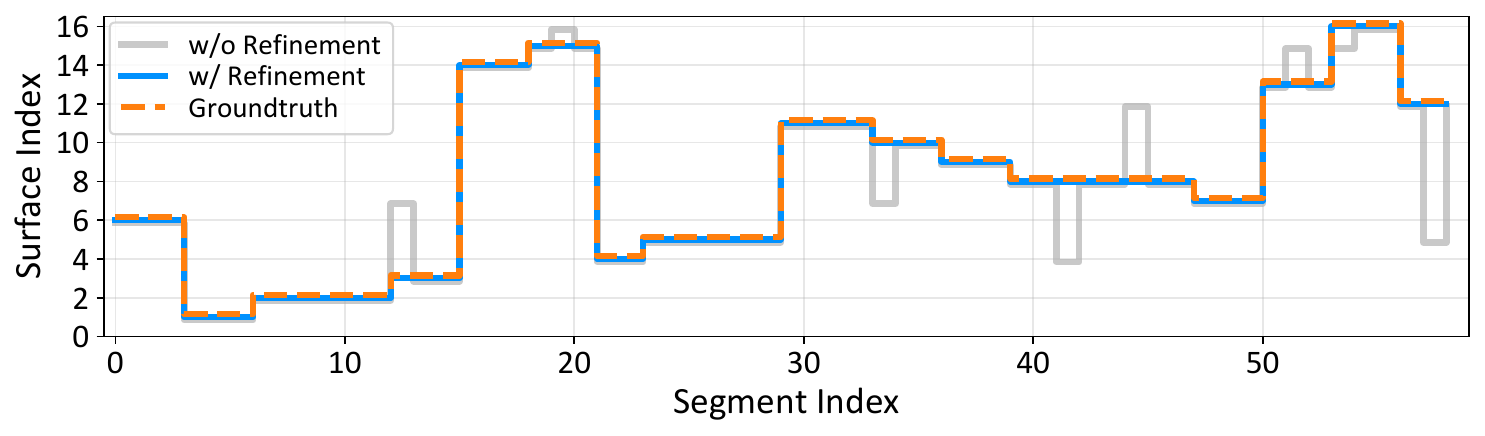}
    \caption{Illustration of Spatio-Temporal Refinement.}
    \label{fig:refinement}
    \vspace{-10pt}
\end{figure}

\subsection{Spatio-Temporal Refinement}
\label{sec:classification.refinement}
\systemname uses the spatio-temporal consistency of brushing sequences to correct the classifier's occasional errors. These errors appear on single strokes as short isolated jumps in an otherwise correct sequence. As users move the brush head smoothly across the teeth, consecutive strokes usually stay on the same surface or move to an adjacent one \cite{oral_seq_survey1_1979,oral_seq_survey2_2015}. An isolated jump breaks this consistency, exposing the misclassified stroke. We therefore model the strokes of one brushing session as a hidden Markov chain and decode it with the Viterbi algorithm \cite{viterbi2003error}. The strokes, in the order they occur, form the time steps of the chain, and the hidden state at each step is the surface being brushed. The observation at each step is the stroke's signal, which the classifier has already summarized into a probability vector $\mathbf{p}$ over the 16 surfaces (Eq.~\ref{eq:fusion}), so we take $\mathbf{p}$ as the emission probabilities of that step.

The transition probabilities come from the layout of the mouth, and the same matrix is used for every user. We describe each surface $s=(r,c,u)$ by its row $r$ (upper or lower), its side $c$ (left, front, or right), and its surface type $u$ (outer, occlusal, or inner). Two surfaces are neighbors if the brush head can move from one to the other without crossing a third surface. For the teeth of the same row and side, the occlusal surface lies between the outer and the inner one, so it is a neighbor of both, and on the front teeth, where there is no occlusal surface in between, the outer and the inner surface are neighbors of each other. Between the two rows on the same side, the surfaces of the same type are neighbors, as the brush head moves from one row to the other without passing any other surface. Ordered from the most to the least likely, a transition either stays on the current surface, moves along the tooth row to the neighboring side while keeping the row and the surface type, moves to another neighbor, or goes anywhere else. Empirically, we set the weights of the four cases to 100, 50, 30, and 5, and normalize the weights of each surface to sum to one. No case receives zero weight, so the refinement penalizes an unusual move but still follows it when the classifier is confident.

The refinement runs once at the end of a session. Starting from a uniform distribution over the 16 surfaces, the algorithm keeps for each stroke and each surface the highest probability among the sequences ending on that surface, then traces the best sequence back from the last stroke. We compute in the log domain to avoid underflow. The cost is $O(16^2 N)$ for $N$ strokes, small compared with the classifier. Fig.~\ref{fig:refinement} shows an example. The raw predictions follow the ground truth on most segments, but occasionally jump to a wrong surface for one or two segments. The refinement removes these isolated jumps, and the refined predictions match the ground truth.

\subsection{Brushing Duration Estimation}
\label{sec:classification.duration}
After refinement, \systemname estimates the brushing duration on each surface by adding up the durations of the segments. The duration of a segment is its length in time, and its surface comes from the refined prediction. Strokes vary in duration, so counting segments does not reflect the brushing time. The accumulated time covers only active brushing, as the idle periods have been removed during segmentation (Sec.~\ref{sec:segmentation}).

%% file: Sections/8_imp.tex
\section{System Implementation}
In this section, we present the implementation of the proposed system, including both hardware and software components. An anonymized copy of the software and hardware is at \url{https://anonymous.4open.science/r/FreeBrush_Artifact/}.

\begin{figure}[t]
\centering
    \begin{subfigure}[t]{\linewidth}
        \centering
        \includegraphics[width=\linewidth]{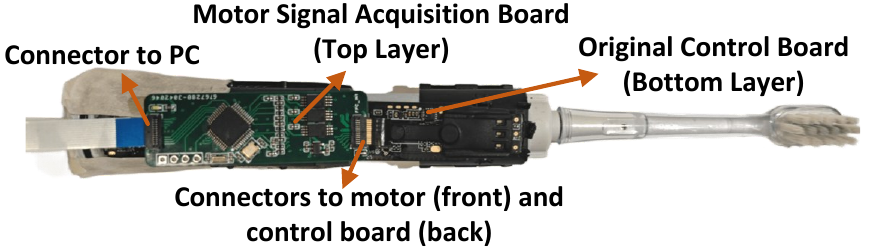}
        \caption{Laifen electric toothbrush integrated with the motor signal acquisition board.}
    \end{subfigure}
    \\[4pt]
    \begin{subfigure}[t]{\linewidth}
        \centering
        \includegraphics[width=\linewidth]{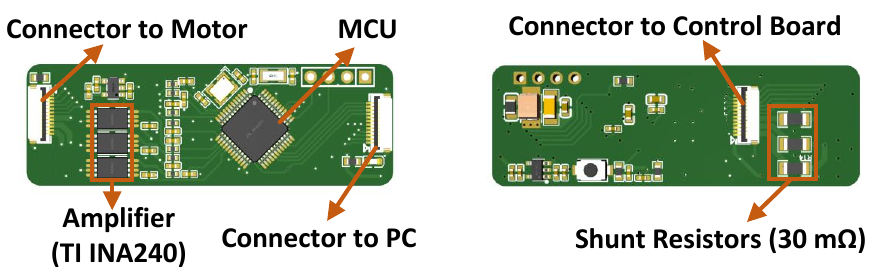}
        \caption{Front and back views of the motor signal acquisition board.}
    \end{subfigure}
    \caption{Data collection hardware prototype.}
    \label{fig:hardware_proto}
    \vspace{-10pt}
\end{figure}

\subsection{Hardware Prototype}
\label{sec:imp.hardware_prototype}
We prototype \systemname\ on a COTS toothbrush (Laifen Wave Electric Toothbrush~\cite{laifen_toothbrush}) together with a small motor signal acquisition board. The toothbrush provides multiple vibration modes and detachable brush heads. We select the gentle mode with a standard gum-care brush head for all experiments unless otherwise specified, as some participants may have sensitive gums and this setting is comfortable for all of them.

The motor signals already exist inside the toothbrush, and reading them is simple. All five channels of motor signals are present on the motor control board, where even an oscilloscope can read them. An oscilloscope is too bulky to carry through a brushing session and cannot log the five channels across many sessions, so we build the motor signal acquisition board to replace it. As shown in Fig.~\ref{fig:hardware_proto}(a), we insert it between the control board and the motor through their flexible printed circuit connectors.

The board digitizes the five signal channels and streams them to a laptop for analysis. As shown in Fig.~\ref{fig:hardware_proto}(b), the three driving currents flow through 30\,m$\Omega$ shunt resistors, whose voltage drops are amplified by TI INA240 amplifiers, and the two position feedback signals are routed directly to ADC inputs. All five channels are digitized by an onboard STM32F103 MCU at 10~kHz per channel and streamed to the laptop over UART. The laptop serves only as the analysis platform in our experiments. The trained system runs in real time on a Raspberry Pi 4B (Sec.~\ref{sec:eval.computational_cost}). The board's material cost is approximately \$6.75.

The board observes the signals without disturbing the toothbrush. The shunts in series with the windings drop only a few millivolts at the toothbrush's sub-ampere driving currents, and the high-impedance ADC inputs draw almost no current from the position feedback lines, so the control board sees the same signals as before.

A manufacturer does not need the board or any additional sensor to deploy the system, as measuring the driving currents and position feedback is necessary for driving the motor in the toothbrush, as described in Sec.~\ref{sec:background.bldc_structure}.

\subsection{Software Implementation}
We implement the classification model and the data processing algorithm in Python with PyTorch 2.6.0. The global model is trained on a workstation running Ubuntu 22.04 with an Intel Xeon Silver 4310 CPU and an NVIDIA GeForce RTX 3090 GPU, which also runs the offline experiments. The learning rate is $10^{-3}$ for training the global model and $10^{-4}$ for fine-tuning, and the batch size is 32. Training runs for up to 100 epochs with early stopping on the validation loss, and fine-tuning runs for 80 epochs. Training the global model takes approximately 33 seconds per epoch, while fine-tuning a personalized model takes approximately 1 second per epoch. The trained system runs in real time on a Raspberry Pi 4B (Sec.~\ref{sec:eval.computational_cost}).

%% file: Sections/9_evaluation.tex
\begin{figure}[t]
    \centering
    \setlength{\abovecaptionskip}{4pt}
    \includegraphics[width=0.6\linewidth]{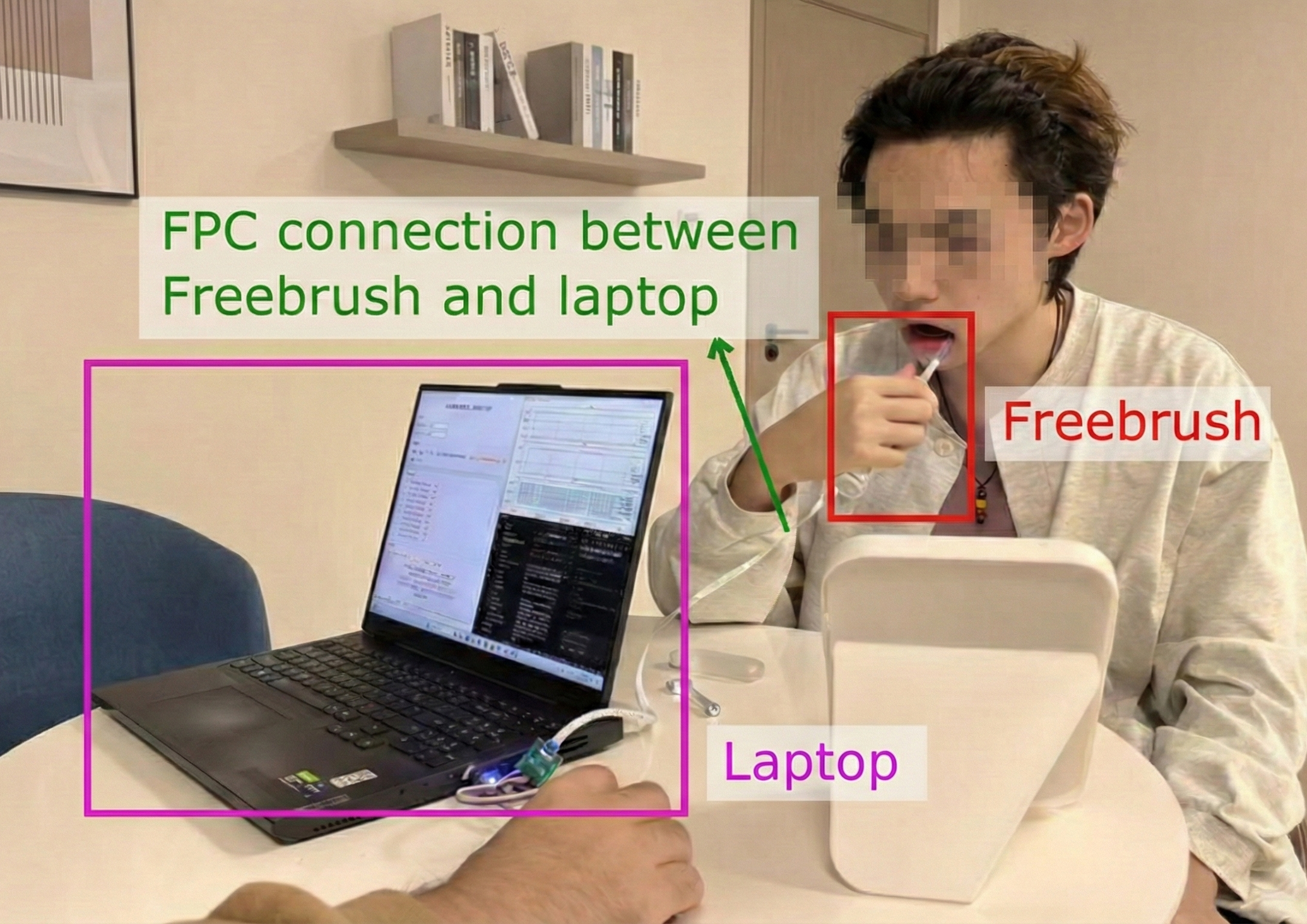}
    \caption{Data-collection setup with the \systemname prototype.}
    \label{fig:eval.experiment_scenario}
\end{figure}

\begin{figure*}[t]
    \centering
    \begin{minipage}[c]{0.42\textwidth}
        \centering
        \includegraphics[width=0.72\linewidth]{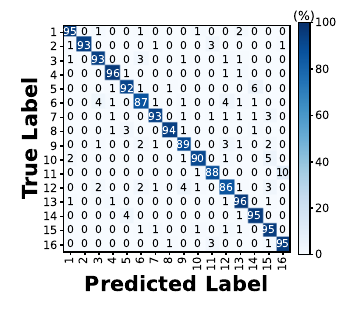}
        \captionof{figure}{Confusion matrix of surface recognition using indices from Fig.~\ref{fig:oral_cavity_property}.}
        \label{fig:eval.rec_confusion}
    \end{minipage}%
    \hfill%
    \begin{minipage}[c]{0.54\textwidth}
        \centering
        \captionof{table}{Comparison of recognition resolution, participant count, and accuracy with prior work.}
        \label{tab:comparison_related_works}
        \small
        \setlength{\tabcolsep}{4pt}
        \renewcommand{\arraystretch}{1.18}
        \begin{tabular}{@{}lccc@{}}
            \toprule
            \textbf{Work} & \textbf{Resolution} & \textbf{\#Part.} & \textbf{Acc.} \\
            \midrule
            SmarTeeth~\cite{yangqiang_smarteeth_2025}
                & 16 surfaces & 13 & 90.3\% \\
            BrushBuds~\cite{yang2024brushbuds}
                & 6 surfaces & 13 & 84.3\% \\
            ToothFairy~\cite{wang_toothfairy_2024}
                & Tooth-by-tooth\textsuperscript{1} & 10 & 90.5\% \\
            LiT~\cite{chen2023lit}
                & 16 surfaces & 16 & 95.3\% \\
            MET~\cite{huang2020met}
                & 15 surfaces & 14 & 85.3\% \\
            mTeeth~\cite{akther_mteeth_2021}
                & 9 classes & 19\textsuperscript{2} & 86.0\% F1\textsuperscript{3} \\
            \textbf{Ours}
                & 16 surfaces & \blue{18} & \blue{92.8\%} \\
            \bottomrule
        \end{tabular}

        \vspace{2pt}
        \begin{minipage}{0.98\linewidth}
            \footnotesize
            \textit{Notes:} \textsuperscript{1}\,ToothFairy identifies each tooth but
            can hardly distinguish inner vs.\ outer surfaces;
            \textsuperscript{2}\,mTeeth annotates 114 brushing episodes from 19
            participants selected from a 30-participant dataset;
            \textsuperscript{3}\,F1-score rather than accuracy.
        \end{minipage}
    \end{minipage}
\end{figure*}

\section{Evaluation}
\label{sec:eval}

\subsection{Evaluation Setup}
\label{sec:eval.setup}
\textbf{Data Collection.}
We recruit \blue{18} participants (\blue{12} males and \blue{6} females, aged \blue{18} to \blue{32}) from the institution community. All participants provided informed consent prior to participation.\footnote{The study was approved by the Institutional Review Board of our institution.} We collect brushing data with our prototype built on a Laifen electric toothbrush (Sec.~\ref{sec:imp.hardware_prototype}). Prior toothbrushing monitoring studies instruct their participants in the Bass brushing technique before data collection~\cite{yangqiang_smarteeth_2025,yang2024brushbuds,wang_toothfairy_2024,chen2023lit}. We follow the same practice, and all participants watch an instructional video to learn the technique before data collection. Then, each participant completes two registration sessions following the protocol in Sec.~\ref{sec:classification.reg}. Each participant completes \blue{10} brushing sessions over 7 days, with each session lasting around 2 minutes. Participants brush with their preferred duration, speed, and force while seated. Mirrors are provided only for participants' reference. We also collect additional sessions to evaluate the system's robustness under different usage scenarios, brush heads, toothpaste types, vibration modes, bristle wear levels, and long-term use.

All data are collected indoors, as shown in Fig.~\ref{fig:eval.experiment_scenario}. Unless otherwise specified, participants use the \blue{standard gum-care brush head} and the gentle vibration mode. They brush without toothpaste to avoid oral irritation during consecutive sessions. In total, we collect \blue{286} brushing sessions and \blue{16,281} active-brushing segments across 16 surface classes. The posterior surfaces contain an average of \blue{852} segments per surface (SD=\blue{11}), while the anterior surfaces contain \blue{1,514} segments per surface (SD=\blue{82}), as each anterior surface covers six teeth while a posterior one covers four.

\textbf{Ground Truth.}
We obtain surface labels without imposing a fixed brushing order. Prior toothbrushing-monitoring studies commonly use guided brushing instructions and fixed orders to ensure reliable labels~\cite{hussain2021you,yangqiang_smarteeth_2025,huang2016toothbrushing,luo2019brush}. Instead, before each session, we ask each participant to report their habitual brushing order, and we collect sessions for every order if a participant reports more than one. The participant then brushes naturally following the reported order, with no required duration for any surface. During brushing, an observer records the start and end time of every surface interval without interacting with the participant. The reported order and these timestamps give the ground-truth surface label and duration of each interval, and the time between two intervals is idle. In total, the dataset covers \blue{36} distinct brushing orders across the participants.
 For each brushing segment produced by the segmentation method (Sec.~\ref{sec:segmentation}), we use the surface label of the annotated interval containing it as the ground truth.

\textbf{Evaluation Protocol.}
We use leave-one-user-out cross-validation unless otherwise specified. In each fold, we hold out one participant for testing, train the global model on the remaining participants, and personalize it for the held-out participant with their two registration sessions (Sec.~\ref{sec:classification.reg}). The same sessions also provide the 16 surface reference vectors. Apart from these two sessions, the system sees no data from the held-out participant before testing.

\textbf{Evaluation Metrics.}
\label{sec:eval.metrics}
We report two metrics, one measuring whether each segment is classified correctly, and the other measuring whether the brushing time on each surface is counted correctly. Given $N$ active-brushing segments with ground-truth labels $y_i$ and predictions $\hat{y}_i$, the \emph{surface recognition accuracy} is
\begin{equation}
A_{\mathrm{surface}}=\frac{1}{N}\sum_{i=1}^{N}\mathbf{1}[\hat{y}_i=y_i].
\end{equation}
For each surface $k$, let $T_k$ be the number of time samples whose ground-truth state is surface $k$. The \emph{duration estimation accuracy} on this surface is
\begin{equation}
A_{\mathrm{duration}}(k)=\frac{1}{T_k}\sum_{t:\,y(t)=k}\mathbf{1}[\hat{y}(t)=y(t)],
\end{equation}
where $y(t)$ and $\hat{y}(t)$ are the ground-truth and predicted states among the 16 surfaces and idle. We report the average over the 16 surfaces. A prediction within 0.2 seconds of an interval boundary is counted as correct if it matches the ground-truth state on either side of that boundary, as the manually recorded timestamps may have small errors.

\begin{figure*}[t]
    \centering
    \begin{minipage}[t]{0.40\textwidth}
        \centering
        \begin{minipage}[t][3.3cm][t]{\linewidth}
            \centering
            \includegraphics[width=\linewidth]{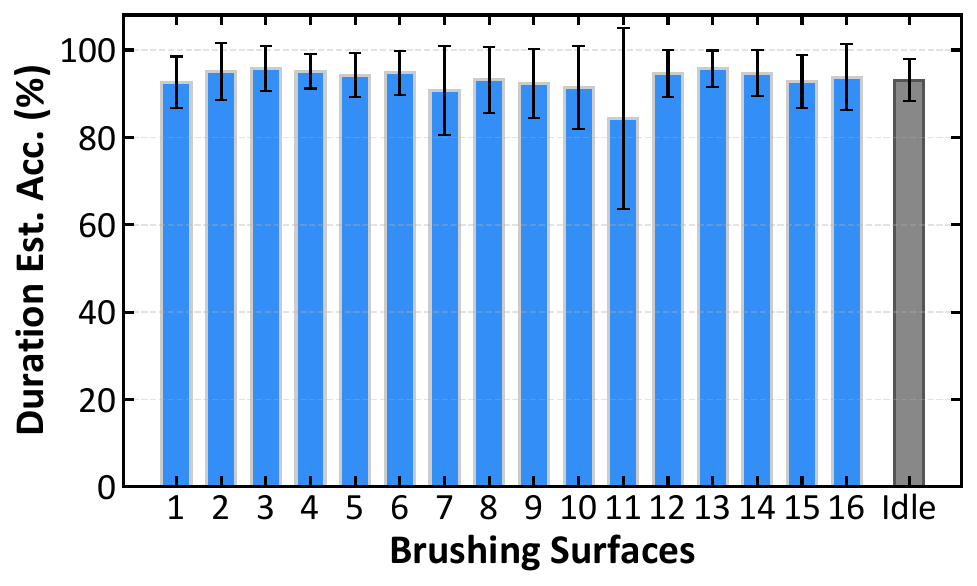}
        \end{minipage}
        \captionof{figure}{Duration estimation accuracy across 16 surfaces (indices in Fig.~\ref{fig:oral_cavity_property}).}
        \label{fig:eval.brushing_duration_accuracy}
    \end{minipage}
    \hfill
    \begin{minipage}[t]{0.28\textwidth}
        \centering
        \begin{minipage}[t][3.3cm][t]{\linewidth}
            \centering
            \includegraphics[width=\linewidth]{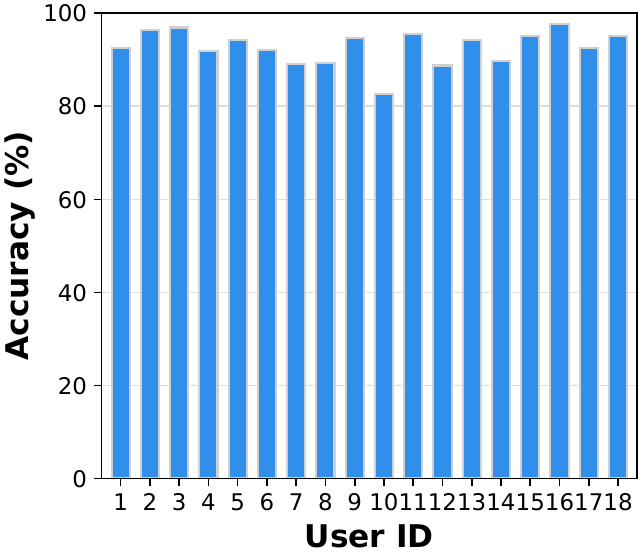}
        \end{minipage}
        \captionof{figure}{Recognition accuracy of each user.}
        \label{fig:eval.user_variance}
    \end{minipage}
    \hfill
    \begin{minipage}[t]{0.28\textwidth}
        \centering
        \begin{minipage}[t][3.3cm][t]{\linewidth}
            \centering
            \includegraphics[width=\linewidth]{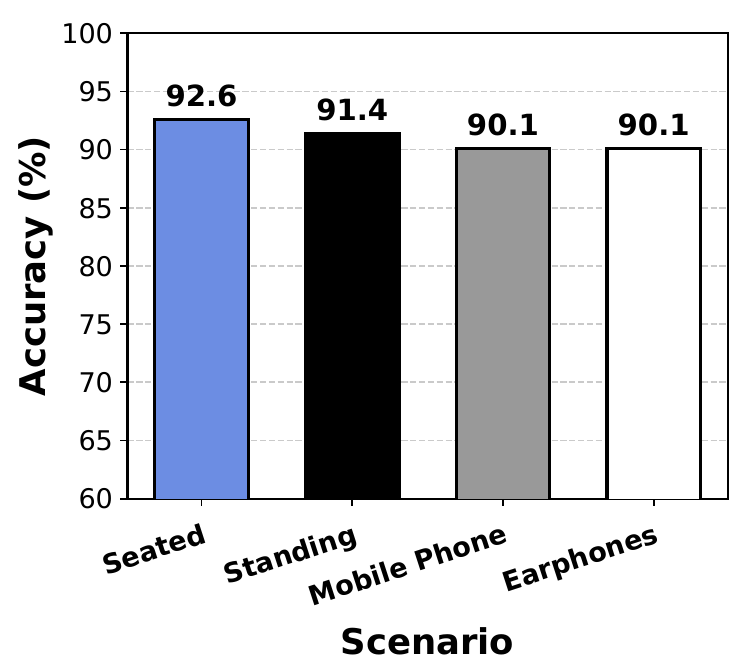}
        \end{minipage}
        \captionof{figure}{Recognition accuracy across four usage scenarios.}
        \label{fig:eval.scenario}
    \end{minipage}
\end{figure*}

\begin{figure*}[t]
    \centering
    \begin{minipage}[t]{0.48\textwidth}
        \centering
        \begin{subfigure}[t]{0.48\linewidth}
            \centering
            \begin{minipage}[t][3.2cm]{\linewidth}
                \centering
                \raisebox{0.4cm}{\includegraphics[height=2.8cm]{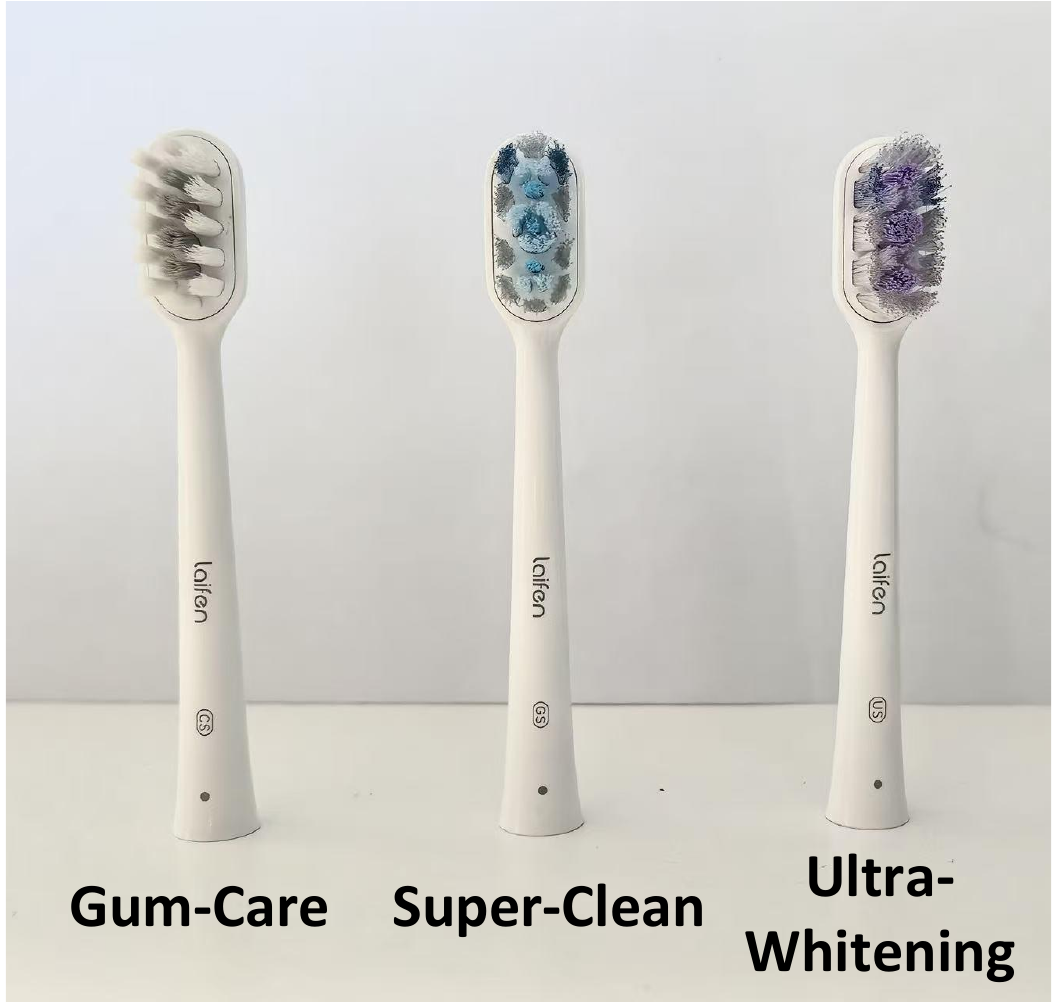}}
            \end{minipage}
            \caption{Three types of brush heads.}
        \end{subfigure}
        \hfill
        \begin{subfigure}[t]{0.48\linewidth}
            \centering
            \begin{minipage}[t][3.2cm]{\linewidth}
                \centering
                \includegraphics[height=3.2cm]{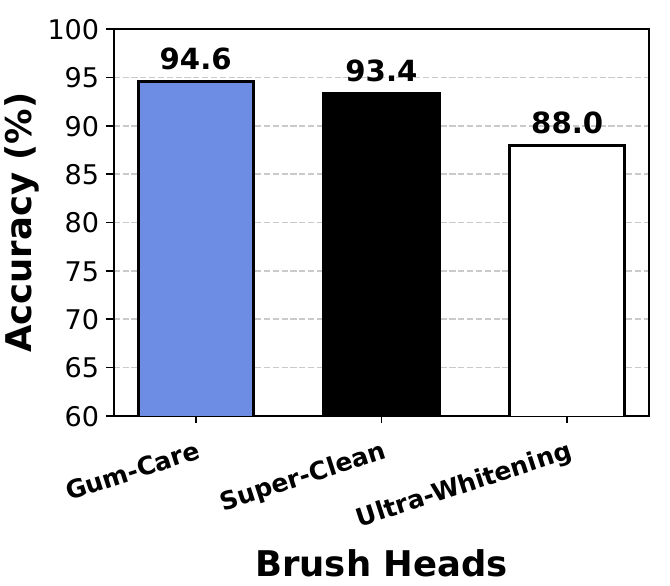}
            \end{minipage}
            \caption{Recognition accuracy with different brush heads.}
        \end{subfigure}
        \addtocounter{figure}{-1}
        \captionof{figure}{Robustness evaluation with different brush heads.}
        \label{fig:eval.brush_head}
    \end{minipage}%
    \hfill%
    \begin{minipage}[t]{0.48\textwidth}
        \centering
        \begin{subfigure}[t]{0.45\linewidth}
            \centering
            \raisebox{0.4cm}{\includegraphics[height=2.55cm,keepaspectratio]{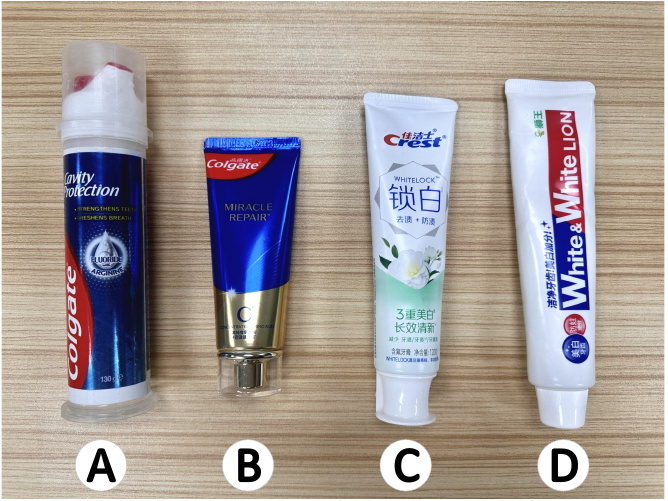}}
            \caption{Toothpastes.}
        \end{subfigure}
        \hfill
        \begin{subfigure}[t]{0.51\linewidth}
            \centering
            \includegraphics[height=2.9cm,keepaspectratio]{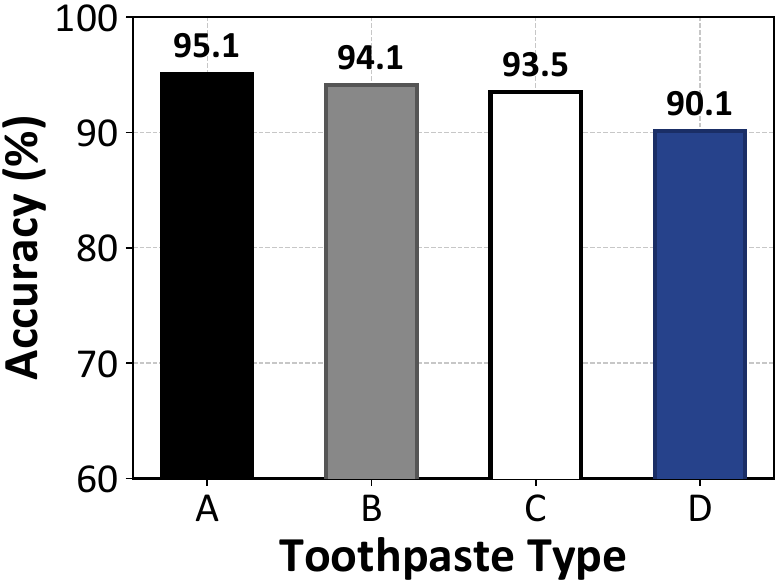}
            \caption{Recognition accuracy.}
        \end{subfigure}
        \captionof{figure}{Robustness evaluation on four different toothpastes.}
        \label{fig:eval.toothpaste}
    \end{minipage}
\end{figure*}

\subsection{Overall Performance}
\label{sec:eval.overall}

\textbf{Surface Recognition.}
\label{sec:eval.overall.surface_recognition}
\systemname achieves \blue{92.8\%} average surface recognition accuracy across 16 tooth surfaces. Fig.~\ref{fig:eval.rec_confusion} shows the confusion matrix. The per-surface accuracies range from \blue{85.6\%} to \blue{96.0\%}, and \blue{12} of the 16 surfaces reach at least 90\%. The errors follow the layout of the mouth instead of spreading over the matrix. Most misclassifications fall on surfaces adjacent to the true one along the tooth row, such as \blue{10\%} of the upper left outer surface (ID~\blue{11}) taken as the upper front outer surface (ID~\blue{16}). Three of the four surfaces below 90\% are the lower left occlusal surface (ID~\blue{6}), the upper right occlusal surface (ID~\blue{9}), and the upper left occlusal surface (ID~\blue{12}), which receive the least constraint from the surrounding soft tissue and are relatively hard to distinguish. However, the system still achieves an accuracy of \blue{85.6\%} or above with these surfaces.

\textbf{Brushing Duration Estimation.}
\label{sec:eval.duration_accuracy_estimation}
Previous experiments demonstrate \systemname's performance in surface recognition. However, accurate estimation of brushing duration is also important for identifying under-brushed and over-brushed surfaces. We therefore evaluate brushing duration estimation across all 16 tooth surfaces, following the duration estimation accuracy defined in Sec.~\ref{sec:eval.metrics}. Fig.~\ref{fig:eval.brushing_duration_accuracy} shows the estimation accuracy for each surface. While the accuracy varies slightly across surfaces, the overall performance remains high, with an average duration estimation accuracy of \blue{92.0\%}. The results show that \systemname can accurately estimate brushing duration for identifying under-brushed and over-brushed surfaces.

\textbf{Comparison with Prior Work.}
\label{sec:eval.overall.comparison}
\systemname recognizes all 16 tooth surfaces with an accuracy comparable to prior work. Table~\ref{tab:comparison_related_works} lists the resolution, the participant count, and the accuracy of representative systems, all of which use sensing modalities different from ours. SmarTeeth and LiT also work at 16 surfaces and report 90.3\% and 95.3\%. The other systems work at different resolutions and report lower accuracies. Our study involves \blue{18} participants, comparable to these systems.

\subsection{Robustness}
We evaluate \systemname across users, usage scenarios, brush heads, toothpaste types, vibration modes, long-term use, and bristle wear. These experiments follow the settings and report the surface recognition accuracy defined in Sec.~\ref{sec:eval.metrics} unless stated otherwise. These robustness experiments involve \blue{5} of the participants and \blue{130} sessions in total, except the \textit{User Variability} experiment, which reuses the data from the overall experiment.

\textbf{User Variability.}
\label{sec:eval.robust.user_variance}
To assess performance variation across users, Fig.~\ref{fig:eval.user_variance} reports each participant's surface-recognition accuracy following the overall experiment setting (Sec.~\ref{sec:eval.overall.surface_recognition}). Thirteen participants achieve accuracy of at least \blue{90\%}, four achieve between 85\% and 90\%, and one achieves between 80\% and 85\%. The accuracies range from 82.6\% to 97.7\%. The lowest comes from User \blue{10}, who wears braces, as the braces may change how the brush head contacts the teeth and disturb the motor load. Every other participant reaches at least 85\%.

\textbf{Impact of Usage Scenarios.}
\label{sec:eval.robust.usage_scenarios}
We evaluate \systemname under four usage scenarios that reflect real-world conditions: (i) seated (default), (ii) standing, (iii) brushing with one hand while using a mobile phone with the other, and (iv) brushing while wearing earphones.
Fig.~\ref{fig:eval.scenario} shows the surface recognition accuracy. The seated, standing, mobile-phone, and earphone scenarios achieve \blue{92.6\%}, 91.4\%, 90.1\%, and 90.1\% accuracy, respectively. The maximum difference is 2.5 percentage points. These scenarios change the user's posture and the use of the other hand, but they leave the oral structures that load the brush head unchanged, so they have limited impact on surface recognition.

\begin{figure*}[t]
    \centering
      \begin{minipage}[t]{0.52\textwidth}
        \centering
        \begin{minipage}[t]{0.48\linewidth}
            \centering
            \includegraphics[width=\linewidth]{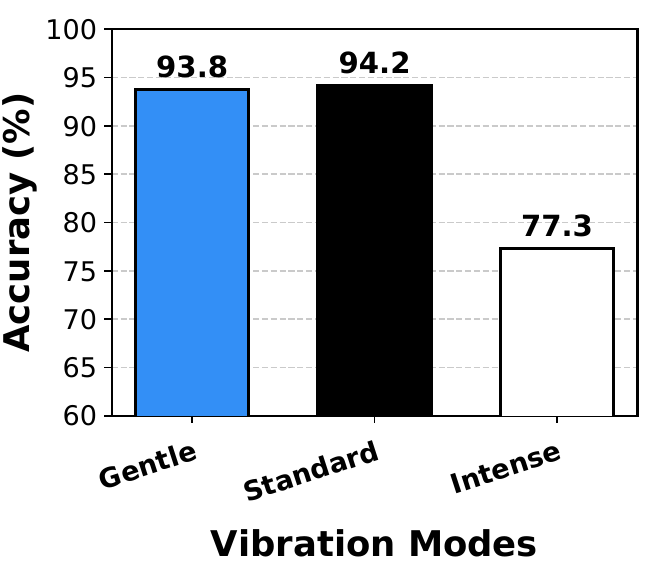}
            \captionof{figure}{Recognition accuracy across vibration modes.}
            \label{fig:eval.vibration_modes}
        \end{minipage}%
        \hfill%
        \begin{minipage}[t]{0.48\linewidth}
            \centering
            \includegraphics[width=\linewidth]{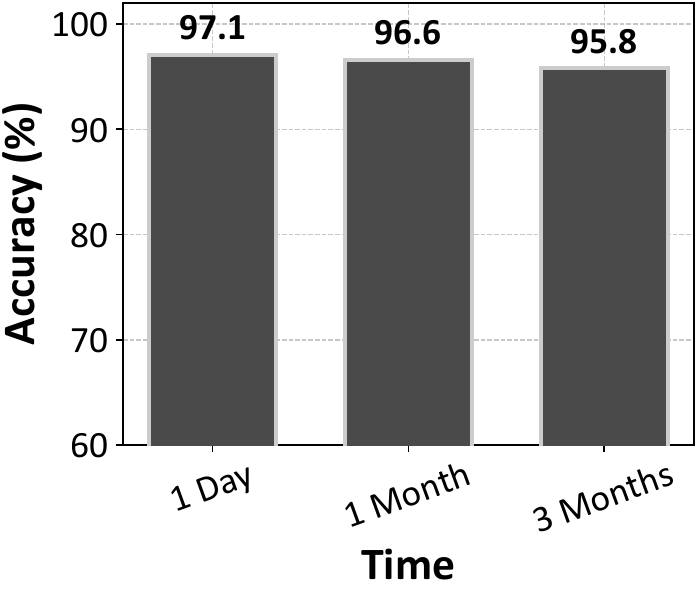}
            \captionof{figure}{Recognition accuracy at 1 day, 1 month, and 3 months after registration.}
            \label{fig:eval.temporal_stability}
        \end{minipage}
    \end{minipage}
    \hfill%
      \begin{minipage}[t]{0.45\textwidth}
        \centering
        \begin{subfigure}[t]{0.42\linewidth}
            \centering
            \includegraphics[width=0.98\linewidth]{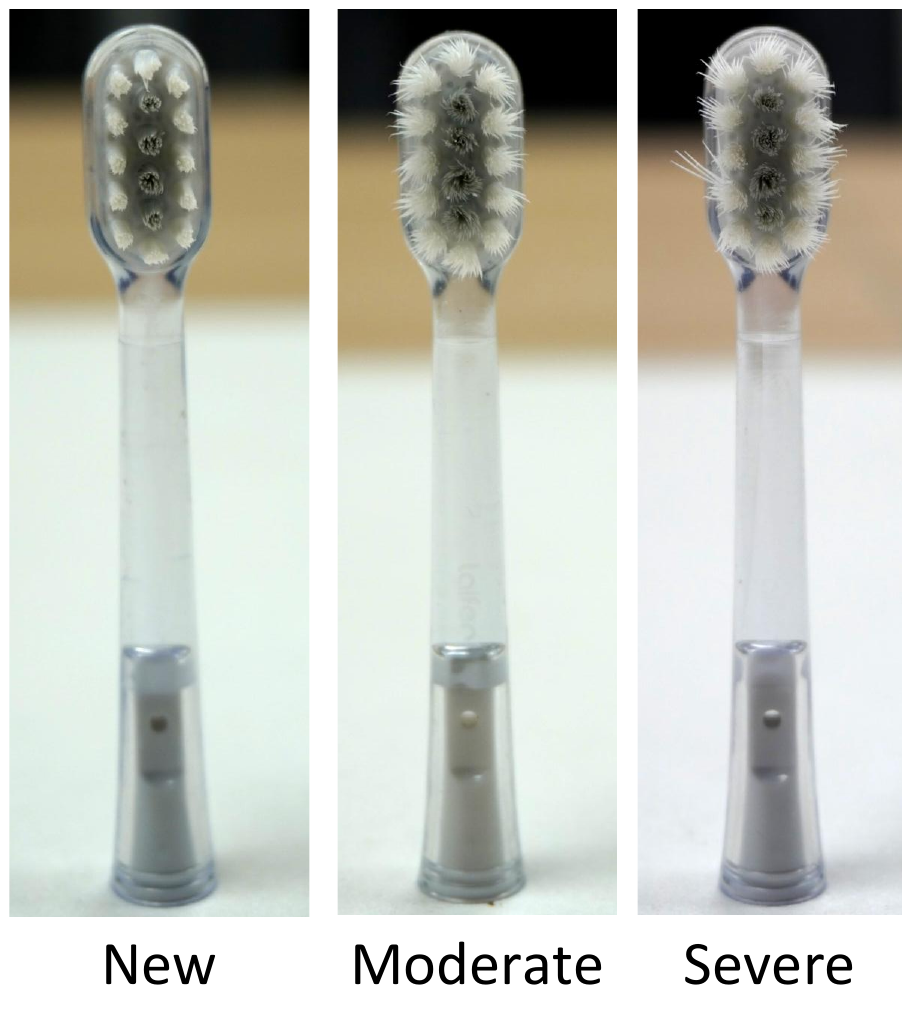}
            \caption{Used brush heads with different wear levels.}
        \end{subfigure}%
        \hfill%
        \begin{subfigure}[t]{0.53\linewidth}
            \centering
            \includegraphics[width=1.05\linewidth]{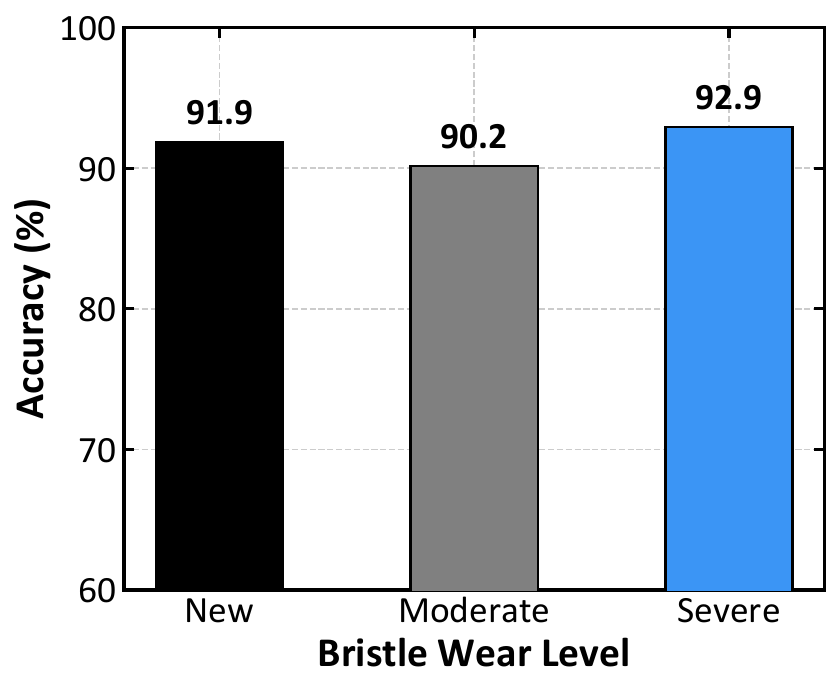}
            \caption{Recognition accuracy across different bristle wear levels.}
        \end{subfigure}
        \addtocounter{figure}{-1}
        \captionof{figure}{Robustness evaluation under different bristle wear levels.}
        \label{fig:eval.bristle_wear}
    \end{minipage}%
\end{figure*}

\begin{figure}[t]
    \centering
    \begin{minipage}[t]{0.48\linewidth}
        \centering
        \includegraphics[width=\linewidth]{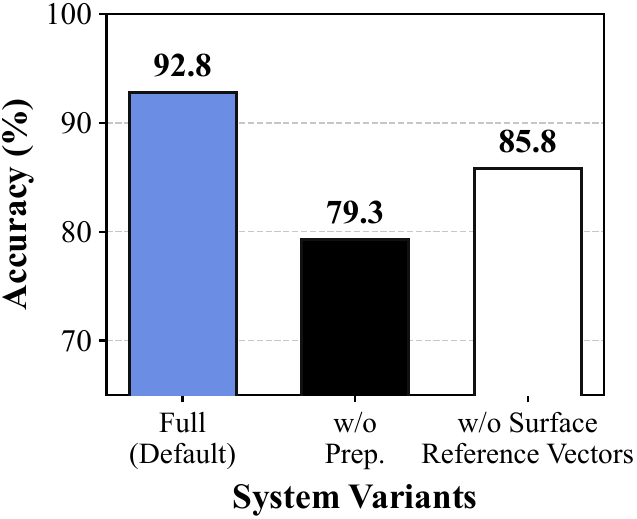}
        \captionof{figure}{Surface-recognition accuracy of the full system and variants without preprocessing or surface reference vectors.}
        \label{fig:eval.ablation_components}
    \end{minipage}%
    \hfill%
    \begin{minipage}[t]{0.48\linewidth}
        \centering
        \includegraphics[width=\linewidth]{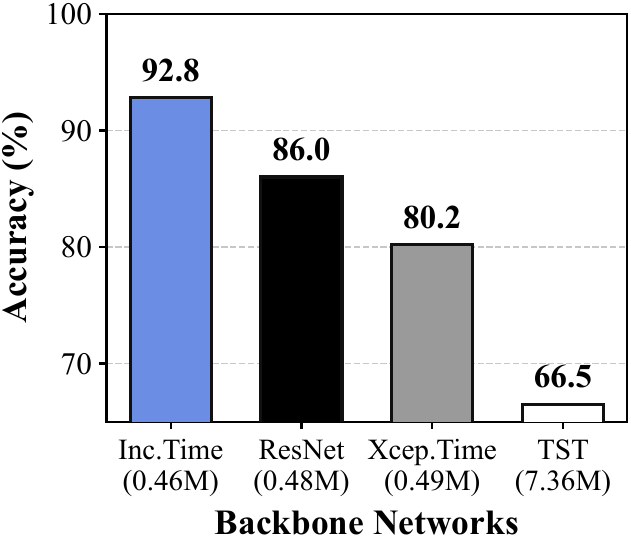}
        \captionof{figure}{Surface-recognition accuracy with four backbone networks. Parameter count is below the name.}
        \label{fig:eval.ablation_backbone}
    \end{minipage}
\end{figure}

\textbf{Impact of Different Brush Heads.}
\label{sec:eval.robust.brush_heads}
COTS electric toothbrushes typically offer multiple brush head options to cater to different user preferences and oral care needs. We evaluate \systemname's performance across three brush head types: (i) standard gum-care head (default), (ii) super-clean head, and (iii) ultra-whitening head. Each brush head has distinct bristle arrangements and stiffness, as shown in Fig.~\ref{fig:eval.brush_head}(a) from left to right, which may influence brushing patterns.

The system is trained on data collected with the standard gum-care head and tested on all three heads. Fig.~\ref{fig:eval.brush_head}(b) shows that the super-clean and ultra-whitening heads achieve surface recognition accuracies of \blue{93.4\%} and \blue{88.0\%}, respectively, compared with \blue{94.6\%} for the standard gum-care head. The maximum difference across the three heads is 6.6 percentage points. Thus, brush-head design affects recognition accuracy, but the system achieves at least 88.0\% across all tested heads.

\textbf{Impact of Toothpaste Types.}
\label{sec:eval.robust.toothpaste}
Prior experiments do not use toothpaste to avoid the long-time stimulation toothpaste on participants' teeth and gums. However, in real-world usage, toothpaste is commonly applied during brushing. Different toothpaste formulations may alter the interaction between the brush head and tooth surfaces, potentially affecting the motor-load feedback patterns captured by our sensing system. We evaluate \systemname's performance across four COTS toothpaste types with varying ingredients and textures: (A) \blue{Colgate Cavity Protection}, (B) \blue{Colgate Miracle Repair}, (C) \blue{Crest Whitelock}, and (D) \blue{Lion White\&White}, as shown in Fig.~\ref{fig:eval.toothpaste}(a). These four toothpastes cover three major functional categories: caries prevention, enamel repair, and whitening. The first three primarily use hydrated silica as the abrasive agent, whereas the last one uses calcium carbonate.

We train the model using data collected without using toothpaste and test it on all four toothpastes. Fig.~\ref{fig:eval.toothpaste}(b) shows the surface recognition accuracy for each type. Toothpastes A, B, and C achieve \blue{95.1\%}, \blue{94.1\%}, and \blue{93.5\%}, respectively. Toothpaste D yields \blue{90.1\%}, showing a slight decrease, likely because it is the only calcium-carbonate-based formulation, whereas the others use hydrated silica. This different abrasive composition may subtly affect motor signal patterns. However, the drop remains limited, and \systemname still achieves an accuracy above \blue{90\%} on all four toothpastes.

\textbf{Impact of Vibration Modes.}
\label{sec:eval.robust.vibration_modes}
COTS electric toothbrushes typically provide multiple vibration modes to accommodate different cleaning preferences. We evaluate \systemname across three commonly used modes: (i) gentle (default), (ii) standard, and (iii) intense. These modes differ in the frequency and the amplitude of the brush head's oscillation. The reference-frame alignment removes the modulation from the rotor's magnetic orientation but keeps the brush head's oscillation in $I_q$ (Sec.~\ref{sec:preprocessing}), so the surface-related patterns are harder to recognize when the mode changes.

The system is trained on data collected in the gentle mode and tested on data from all three modes. As shown in Fig.~\ref{fig:eval.vibration_modes}, the standard mode achieves a surface recognition accuracy of \blue{94.2\%}, 0.4 percentage points higher than the gentle mode (\blue{93.8\%}), whereas the intense mode drops to \blue{77.3\%} without additional adaptation. The intense mode differs the most from the gentle mode in oscillation amplitude and frequency, and the motor supplies more current to sustain a larger and faster swing, so both the level of $I_q$ and the shape of its swing move furthest from the data the model is trained on. With one additional fine-tuning session, the accuracy in the intense mode improves to \blue{89.7\%}. We use the gentle mode as the default in our primary experiments to reduce potential gum discomfort and accommodate variations in participants' oral sensitivity.

\textbf{Long-term Use.}
\label{sec:eval.long_term_use}
\systemname requires only a one-time registration, so it should remain reliable as users continue using the toothbrush over time. We test the system at 1 day, 1 month, and 3 months after registration without any re-registration. As shown in Fig.~\ref{fig:eval.temporal_stability}, the surface recognition accuracies at 1 month and 3 months are \blue{96.6\%} and \blue{95.8\%}, compared to \blue{97.1\%} at 1 day. The surface-related patterns are governed by the oral structure, which stays nearly unchanged over months, so the registered reference vectors remain valid. The small drop across the three months falls within the system's inherent error. These results show that \systemname remains stable over time and does not require repeated registration.

\begin{figure}[t]
    \centering
    \begin{subfigure}[t]{0.48\linewidth}
        \centering
        \includegraphics[width=0.95\linewidth]{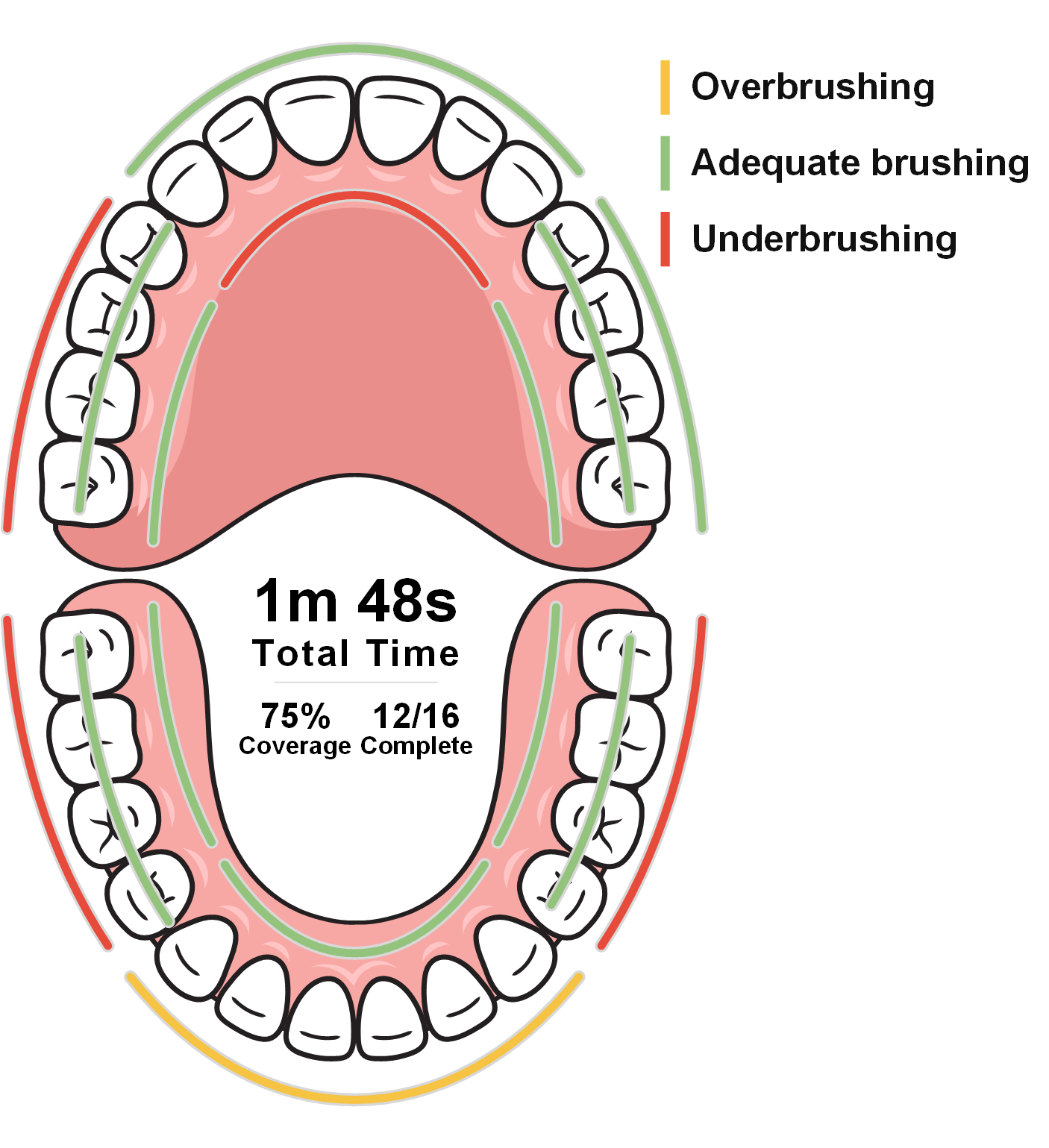}
        \caption{Initial brushing result. The under-brushed surfaces are highlighted in red.}
    \end{subfigure}%
    \hfill%
    \begin{subfigure}[t]{0.48\linewidth}
        \centering
        \includegraphics[width=0.95\linewidth]{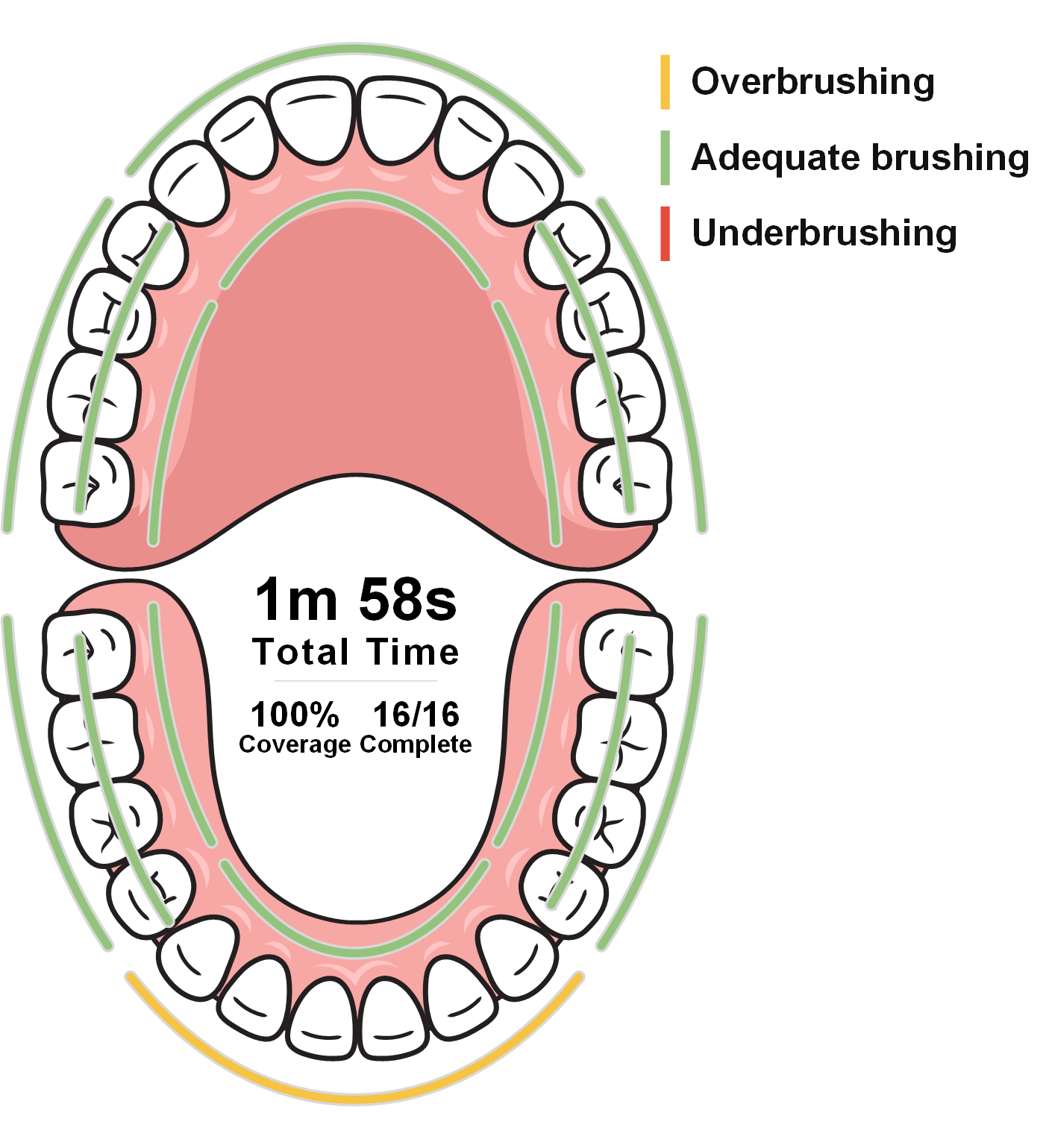}
        \caption{After re-brushing the under-brushed surfaces based on \systemname's feedback.}
    \end{subfigure}
    \caption{Case study of brushing improvement based on \systemname's feedback.}
    \label{fig:eval.feedback_case}
\end{figure}

\textbf{Bristle Wear.}
\label{sec:eval.robust.bristle_wear}
We also evaluate the impact of bristle wear on \systemname's performance. We collect data from 3 brush heads at different wear levels: new, moderately worn, and severely worn, as shown in Fig.~\ref{fig:eval.bristle_wear}(a). The new brush head has intact and regularly arranged bristles, the moderately worn head shows visible bristle spreading, and the severely worn head has noticeably splayed and deformed bristles. The system is trained on data from the new brush head and tested on all three heads. Fig.~\ref{fig:eval.bristle_wear}(b) shows that the moderately worn head achieves a surface recognition accuracy of \blue{90.2\%}, while the severely worn head achieves \blue{92.9\%}, compared to \blue{91.9\%} for the new head. The three accuracies stay within 2.7 percentage points of each other and do not follow the wear level. The surface-related patterns are governed by the oral structure rather than by the bristle condition, so bristle wear has little effect on surface recognition.

\subsection{Ablation Study}
\label{sec:eval.ablation}
The ablation study examines the contributions of preprocessing and the surface reference vectors. We remove each component while retaining the remaining pipeline, and we separately replace InceptionTime with three time-series backbones.

\textbf{Impact of System Components.}
To examine these two components under the overall evaluation setting, we first remove the complete preprocessing block, so the five channels of raw motor signals are fed directly to the model, still modulated by the brush head's oscillation and without any dimension reduction. Segmentation and the personalized classifier remain unchanged. As shown in Fig.~\ref{fig:eval.ablation_components}, the accuracy decreases from \blue{92.8\%} to \blue{79.3\%}. We then retain preprocessing and user-specific fine-tuning but remove the surface reference vectors, leaving the fine-tuned classification head to make the decision. This variant reaches \blue{85.8\%}. The two results show that both preprocessing and surface reference-vector comparison contribute to the final accuracy.

\textbf{Impact of Backbone Network.}
To verify the choice of InceptionTime as the backbone, we replace the latent layers with ResNet~\cite{wang2017timeseries}, XceptionTime~\cite{rahimian2020xceptiontime}, and the Time Series Transformer (TST)~\cite{zerveas2021transformer}, each with the default configuration in its original paper, and keep the rest of the pipeline unchanged. Fig.~\ref{fig:eval.ablation_backbone} compares InceptionTime, ResNet, XceptionTime, and the Time Series Transformer (TST), with each model's parameter count shown below its name. ResNet and XceptionTime are similar in size to InceptionTime but reach \blue{86.0\%} and \blue{80.2\%}. The Time Series Transformer is over ten times larger yet reaches only \blue{66.5\%}, as a transformer of this size needs far more training data than our dataset provides. InceptionTime achieves the highest accuracy with a small parameter count.

\subsection{Case Study on Brushing Feedback}
\label{sec:eval.closed_loop_feedback}

\systemname reports the brushing duration of each surface, and the report tells the user which surfaces to brush more and which to brush less. We visualize the brushing record of each session and highlight under-brushed, over-brushed, and adequately brushed surfaces based on the estimated brushing duration, as shown in Fig.~\ref{fig:eval.feedback_case}. Following \cite{huang2020met}, we label surfaces brushed for less than \blue{5 seconds} as under-brushed, surfaces brushed for more than \blue{10 seconds} as over-brushed, and surfaces brushed between these two thresholds as adequately brushed. We present a case study on one participant to show the report and how the participant acts on it.

In the case study, the participant first brushes with the standard gum-care head and toothpaste A in Sec.~\ref{sec:eval.robust.toothpaste}, using their preferred brushing order, speed, and force. After the initial brushing, \systemname analyzes the brushing record and highlights under-brushed and over-brushed surfaces, shown in red and yellow in Fig.~\ref{fig:eval.feedback_case}(a), respectively. In the initial result, 12 of the 16 surfaces are brushed long enough, one of which is over-brushed, and the remaining 4 are under-brushed. The participant then re-brushes the 4 under-brushed surfaces and is reminded to spend less time on the over-brushed one. As shown in Fig.~\ref{fig:eval.feedback_case}(b), all 16 surfaces reach a sufficient duration after re-brushing the 4 under-brushed surfaces. The case study shows that \systemname's feedback lets the user correct uneven brushing within a single session, helping users build balanced brushing habits.

\subsection{Computational Cost}
\label{sec:eval.computational_cost}
Previous experiments are conducted on powerful desktop-class machines, while real deployments may run on embedded processors with limited resources. To evaluate how the system performs on resource-limited hardware, we benchmark the system on a Raspberry Pi 4B.

The algorithm operates at three time scales. Preprocessing~(Sec.~\ref{sec:preprocessing}) and segmentation~(Sec.~\ref{sec:segmentation}) handle 1-second raw windows and produce variable-length segments (0.8-8~s). Model inference (Sec.~\ref{sec:classification.classifier}) runs once per segment. Spatio-temporal refinement (Sec.~\ref{sec:classification.refinement}) aggregates all predicted segments at the end of a session. The measured latencies are: 28.69 ms for preprocessing and segmentation per 1-second window, 183.41 ms for model inference per segment, and 11.94 ms for spatio-temporal refinement per session. These results show that the full processing algorithm can run efficiently on an edge platform, indicating feasibility for deployment on resource-constrained devices.

%% file: Sections/10_related.tex
\section{Related Work}
Existing toothbrushing-monitoring systems infer brushing locations using camera, photosensor, inertial, magnetic, or acoustic signals. We review them by sensing modality.

\textbf{Camera-based Toothbrushing Monitoring.}
The camera-based approaches visually track the toothbrush during brushing. Playful Toothbrush~\cite{chang_playful_2008} uses an external camera to track an LED-coded toothbrush extension and recognizes brushing across 24 areas. However, camera-based monitoring requires dedicated camera hardware and captures images in a private setting. In contrast, \systemname requires no camera setup and avoids the privacy concerns associated with visual monitoring.

\textbf{Photosensor-based Toothbrushing Monitoring.}
The photosensor-based approaches infer toothbrush position from changes in received light. LiT~\cite{chen2023lit} mounts two photosensors on a commercial LED toothbrush to recognize 16 tooth surfaces. Although LiT provides the same spatial resolution as \systemname, it requires a lighted toothbrush equipped with photosensors. In contrast, \systemname achieves comparable 16-surface recognition without external sensors.

\textbf{IMU-based Toothbrushing Monitoring.}
The IMU-based approaches use accelerometers, gyroscopes, and magnetometers to characterize brushing motion. Hussain et al.~\cite{hussain2021you} use an IMU to recognize 16 brushing activities. Other systems use wrist-worn IMUs to recognize brushing surfaces or motions~\cite{huang2016toothbrushing,luo2019brush,akther_mteeth_2021,feng_hygiea_2024}, while BrushBuds~\cite{yang2024brushbuds} uses IMU signals from earbuds to recognize six surfaces. However, these approaches require sensors attached to the toothbrush or devices worn on the wrist or ears, increasing user burden during brushing. They can also be affected by device placement and unrelated body motion. In contrast, \systemname recognizes 16 surfaces without requiring users to attach or wear an IMU.

\textbf{Magnetic-based Toothbrushing Monitoring.}
The magnetic approaches estimate toothbrush position from changes in a magnetic field. MET~\cite{huang2020met} deploys a magneto-inductive sensor array near the user and recognizes 15 tooth surfaces. However, the sensor array must be installed in the brushing environment. In contrast, \systemname provides finer 16-surface recognition without deploying external sensors.

\textbf{Acoustic-based Toothbrushing Monitoring.}
The acoustic approaches infer brushing activities from sounds transmitted through air or the human body. Smartphone-based methods analyze brushing sounds to recognize brushing actions and estimate brushing performance~\cite{korpela2015evaluating}. Ouyang et al.~\cite{ouyang2017asymmetrical} combine a throat microphone with an earphone to recognize brushed tooth sections. ToothFairy~\cite{wang_toothfairy_2024} uses bone-conducted brushing sounds captured by wired earphones to identify individual teeth, but cannot distinguish the inner and outer surfaces of the same tooth. SmarTeeth~\cite{yangqiang_smarteeth_2025} uses in-ear microphones to recognize 16 tooth surfaces, while Cui et al.~\cite{cui2024dentists} combine throat microphones with a Bluetooth earphone to recognize brushing areas. Although these approaches leave the toothbrush unchanged, they require a nearby recording device or equipment worn around the ear or neck. In contrast, \systemname provides comparable 16-surface recognition without requiring users to wear or deploy any additional device, thereby reducing user burden during brushing. Table~\ref{tab:comparison_related_works} provides a quantitative comparison with representative systems.

%% file: Sections/11_discussion.tex
\section{Discussion}
We discuss the scope and future directions of \systemname in the following:

\textbf{User Registration.}
\systemname uses a one-time registration step to adapt the model to each user, as oral structures differ across users and shift the signals of the same surface. \systemname requires only 2 sessions ($\approx$4 mins) for registration, whereas representative approaches often require 3 or more sessions, or a larger set of labeled brushing episodes, to personalize the model~\cite{akther_mteeth_2021,yangqiang_smarteeth_2025,yang2024brushbuds}. Further research may focus on learning user-invariant representations to reduce per-user registration.

\textbf{System Overhead.}
The signals \systemname needs are already measured inside the electric toothbrush. The toothbrush itself has to read the driving currents and the position feedback to maintain a stable oscillation of the brush head (Sec.~\ref{sec:background.bldc_structure}), so a manufacturer takes them from its own control board and adds no sensor. The motor signal acquisition board in our prototype only taps these signals out of an off-the-shelf product for data collection, and a product would not carry it. The 10~kHz sampling rate is within reach of an entry-level microcontroller that costs less than \blue{\$2}~\cite{stm32f103_price,stm32f103x8_datasheet}. Fine-tuning happens once at registration and can run on the user's phone. Everything after registration runs in real time on a Raspberry Pi 4B (Sec.~\ref{sec:eval.computational_cost}).

\textbf{Toothbrushes with Other Motor Types.}
\systemname is designed for BLDC motor-driven electric toothbrushes, which are becoming increasingly prevalent~\cite{laifen_toothbrush, usmile_y20pro_toothbrush, xiaomi_oscillation_toothbrush}. The system uses the driving currents and the position feedback that BLDC motor control requires (Sec.~\ref{sec:background.bldc_structure}), rather than anything specific to one product, so it carries over across BLDC toothbrushes. We have captured the same signals on BLDC toothbrushes from other manufacturers. Toothbrushes driven by other motor types, such as the linear resonant actuators used in sonic toothbrushes, drive a single coil at its resonance. The load on the brush head still shows in the coil current, as it damps the resonance, so extending \systemname to them is a natural next step, and we leave the study of their signals and preprocessing to future work.

\textbf{Brushing Techniques.}
\systemname is evaluated under the Bass technique, which places the bristles at $45^\circ$ to the gumline and cleans with short strokes. Most dental associations recommend this method~\cite{wainwright2014analysis}, and prior toothbrushing monitoring studies adopt it as well~\cite{yangqiang_smarteeth_2025,yang2024brushbuds,wang_toothfairy_2024,chen2023lit}. Recognizing other brushing techniques would require data collected under those techniques, and we leave it as a next step.

\textbf{Studies on Larger Populations.}
Although our study includes 18 participants, which is comparable to or larger than the participant pools used in representative toothbrushing monitoring studies~\cite{yangqiang_smarteeth_2025,yang2024brushbuds,wang_toothfairy_2024,chen2023lit,huang2020met}, future studies with larger and more diverse populations would further support the broad deployment of \systemname.

%% file: Sections/12_conclusion.tex
\section{Conclusion}
In this work, we propose \systemname, a toothbrushing monitoring system that uses the intrinsic motor signals of electric toothbrushes without any external sensors. When brushing different tooth surfaces, the brush head experiences different constraints from the surrounding oral cavity, and the constraints change the motor's driving currents and position feedback. \systemname uses these signals to recognize the brushed surface among 16 surfaces and estimate the brushing duration on each surface. We prototype the system on a commercial off-the-shelf electric toothbrush and evaluate it with 18 participants. \systemname achieves \blue{92.8\%} accuracy for 16-surface recognition and \blue{92.0\%} accuracy for brushing duration estimation. It remains robust across different toothpastes, brush heads, vibration modes, usage scenarios, and long-term use, and runs on edge devices. With the recognized surfaces and estimated durations, \systemname helps users find under-brushed and over-brushed surfaces and build balanced brushing habits.